\documentstyle[draft,psfig]{mn}
\def\gs{\mathrel{\raise0.35ex\hbox{$\scriptstyle >$}\kern-0.6em 
\lower0.40ex\hbox{{$\scriptstyle \sim$}}}}
\def\ls{\mathrel{\raise0.35ex\hbox{$\scriptstyle <$}\kern-0.6em 
\lower0.40ex\hbox{{$\scriptstyle \sim$}}}}
\def\UB{\hbox{$(U-B)$}}
\def\BI{\hbox{$(B-I)$}}

\begin{document}

\title[The Galaxy Populations of Distant X-ray Clusters]{A Statistical Analysis of the Galaxy Populations of
Distant Luminous X-ray Clusters\footnotemark}

\author[Smail et al.]
{Ian Smail,$^{\! 1}$ Alastair C.\ Edge,$^{\! 2}$
Richard S.\ Ellis$^{2}$ \& Roger D.\ Blandford$^{3}$\\
$^1$ Department of Physics, University of Durham, Durham DH1 3LE \\
$^2$ Institute of Astronomy, Madingley Road, Cambridge CB3 0HA \\
$^3$ Theoretical Astrophysics, Caltech 130-33, Pasadena CA 91125}

\date{\fbox{\sc Accepted Version}}

\label{firstpage}

\maketitle

\begin{abstract}
We present a deep, multi-colour ($UBI$) CCD survey using the Palomar
5-m telescope of a sample of high X-ray luminosity, distant clusters
selected from the ROSAT All-Sky Survey.  The 10 clusters lie in
redshift range $z=0.22$--0.28, an era where evolutionary effects have
been reported in the properties of cluster galaxy populations.  Our
clusters thus provide a well-defined sample of the most massive systems
at these redshifts to quantify the extent and variability of these
evolutionary effects.  The relatively low redshifts of these clusters
also means that simple connections can be made between the galaxy
populations of these clusters and their immediate descendents, local
rich clusters.  Moreover, by concentrating on a narrow redshift range,
we can take advantage of the homogeneity of our cluster sample to
combine the galaxy catalogues from all the clusters to 
analyse statistically the bulk properties of their populations.  We present an
analysis of the cluster galaxy populations using our multi-colour data
to probe the distribution, luminosities and star-formation histories of
galaxies in these regions.  Our aim is to chart the characteristics of
the galaxy populations of massive intermediate redshift clusters and to
combine these into a wider scheme for galaxy evolution in high density
environments.  The core regions of clusters in our sample contain only
a small proportion of star-forming galaxies, and they therefore do not
exhibit a classical ``Butcher-Oemler'' effect.  Focusing on the redder
cluster galaxies we find that their integrated luminosity is well
correlated with the cluster's X-ray temperatures, and hence with
cluster mass.  Furthermore, the typical restframe UV--optical colours
of the luminous elliptical sequences in the clusters exhibit a
remarkably small cluster-to-cluster scatter, $\ls 2$\%, indicating that
these galaxies are highly homogeneous between cluster environments.
However, at fainter magnitudes we observe a marked increase in the
range of mid--UV colours of galaxies possessing strong 4000\AA\ breaks,
as determined from photometry in $\sim 7.5 h^{-1}$ kpc diameter
apertures.  In the light of the apparent decline in the population of
S0 galaxies seen in distant, $z\gs 0.4$, clusters (Dressler et
al.\ 1997), and in view of the luminosities and colours of this
population, we propose that they may be the progenitors of the dominant
S0 population of local rich clusters, caught in the final stage before
they become completely quiescent.  Further studies of this population
will provide a necessary link to connect the evolution observed in
cluster populations at high redshift with the nature of the final
remnants locally.  Observations in the restframe UV will be important
in these studies owing to the relative ease of detecting the signature
of previous activity in this spectral region.  
\end{abstract}

\begin{keywords} cosmology: observations -- clusters --
galaxy evolution -- galaxies: photometry -- galaxies: luminosity function
-- X-ray astronomy.
\end{keywords}

\section{Introduction}

\footnotetext{Based on observations obtained at
Palomar Observatory, which is owned and operated by the California
Institute of Technology.} 

The study of galaxies in rich clusters at earlier epochs has long been
seen as one of the best routes to understanding the role of environment
in the evolution of galaxy populations, as well as more general issues
of galaxy formation.  For this reason a number of groups have studied
the photometric and spectroscopic properties of galaxies in distant
clusters (e.g.\ Butcher \& Oemler 1978, 1984; Couch \& Sharples 1987;
Dressler \& Gunn 1992; Arag\'on-Salamanca et al.\ 1993; Barger et
al.\ 1996).  One of the most interesting conclusions of these studies
has been the realisation that the luminous elliptical population which
dominates local cluster cores (Dressler 1980) has been in place for a
considerable time, at least prior to $z\sim 0.6$ (Smail et al.\ 1997a;
Ellis et al.\ 1997) and possibly earlier than $z\sim 1$ (Faber et
al.\ 1997; Lubin et al.\ 1997).  All indications are that this
population has undergone only passive evolution in the recent past (Van
Dokkum \& Franx 1996; Ellis et al.\ 1997; Barger et al.\ 1997).  In
contrast to the relative stability of the elliptical population, the
other major component of the local cluster population, S0 galaxies, are
claimed to be relatively rare in distant clusters (Dressler et
al.\ 1997).  The recent formation or transformation of this population
and the nature of the process responsible for it are therefore of
considerable interest for understanding the extent of environmental
influences on galaxy morphology.
 
The most dramatic evidence for evolutionary change in the cluster
population, however, is seen in the blue galaxy populations of these
regions (Butcher \& Oemler 1978, 1984; Couch \& Sharples 1987; Dressler
\& Gunn 1992).  Butcher \& Oemler's extensive study of a heterogenous
sample of distant clusters indicated a substantial increase in the blue
populations of rich clusters at $z\gs 0.2$.   Recent HST work on these
systems has shown that these objects are mainly disk galaxies, some of
which are interacting (Couch et al.\ 1994, 1997a, 1997b; Dressler et al.\ 1994;
Smail et al.\ 1997a).  The absence of this population from similar
environments in local clusters may connect with the large present-day
population of S0s in these regions.  Mechanisms proposed for removing
or transforming these star-forming galaxies from the centres of
clusters  include ram-pressure stripping from the ICM (Gunn \& Gott
1972), tidal effects due to the cluster potential (Byrd \& Valtonen
1990) and interactions with other cluster galaxies (Moore et
al.\ 1996).  To date, however, few well-defined samples of distant
clusters have been available to test the different correlations
predicted by these mechanisms (c.f.\ the $L_X$--$f_{\rm sp}$ relation
shown by Edge \& Stewart (1991) or the mass--$f_{\rm sp}$ relation
of Smail et al.\ 1997b).  Such samples are also necessary to
link the observations at different redshifts, by providing
an understanding of the parallel evolution of the structures which the
galaxies inhabit.

Here we describe a statistical analysis of the galaxy populations in a
large, well-defined sample of 10 distant clusters at $z=0.2$--0.3,
selected on the basis of their X-ray emission.  Our sample focuses on
the highest $L_X$ systems, from this we hope to understand the
variation in the properties of massive clusters across a relatively
narrow mass range, and the effects this has had on the galaxy
populations in such regions.  In particular, we will concentrate on the
variation in the star-formation histories of the luminous elliptical
populations between clusters, as determined from their UV--optical
colours.  This is a technique previously only applied to small samples
of 2--3 clusters (Bower, Lucey \& Ellis 1992, BLE; Ellis et al.\ 1997),
here we can compare a homogeneous sample of 10 clusters imaged in
identical conditions to robustly determine the cluster to cluster
scatter.  In particular, we illustrate how our combination of $UBI$
imaging is particularly useful for searching for the signature of
recent star-formation in more quiescent systems.  We show the
advantages of working in the restframe mid--UV for detecting even small
traces of past activity in certain populations (Dorman et al.\ 1995).
We also study the blue populations of these clusters, using the large
positive K correction in $U$ for these galaxies to identify them
relative to the passive cluster galaxies.   We focus on
the varying proportion of this blue, star-forming population in the
clusters and the dependence of this component on the global properties
of the clusters.  Our final aim is to understand the relationships
between the various galactic components of the clusters within the
framework of a simple model for their growth and evolution.

This study is particularly timely given recent theoretical work in this
area (Bower 1991; Moore et al.\ 1996; Baugh et al.\ 1996, 1997).  The
integration of spectral evolution models with simple dynamical models
for structure formation, allow us for the first time to begin to
compare theory and observations in this field (e.g.\ Baugh et
al.\ 1997).  Detailed N-body simulations are also now well placed
to follow the dynamical evolution of galaxies within cluster potentials
(e.g.\ Moore et al.\ 1996, 1997).  It is these types of quantitative
comparisons which will benefit from the well-defined
cluster sample we use.  This paper is thus part of an on-going program
to map out the properties of the galaxy populations in the most massive
collapsed structures in the universe as a function of epoch out to
$z\sim 0.5$.  The data described here deals with the statistical
analysis of the galaxy populations in  luminous X-ray clusters in the
range $0.2\leq z \leq 0.3$.  A similar study using very wide field CCD
imaging and spectroscopy of a sample of luminous X-ray clusters at
$0.07\leq z \leq 0.15$ is underway (O'Hely et al., in prep), this
will provide continuous coverage of the evolution of the most massive
clusters and their galaxy populations from $z\sim 0$--0.3.

Previous work linking the galaxy population of distant clusters with
their global characteristics has concentrated on the blue, star-forming
or ``Butcher-Oemler'' galaxies in distant clusters (Lea \& Henry 1988)
and has found little evidence relating their characteristics to the
X-ray properties of the clusters (although see Wang \& Ulmer 1997).
Nevertheless, there is still much discussion of the role of a cluster's
X-ray-emitting gas in effecting the properties of its galaxy populations.
We note, however, that the necessity for such processes to power the
Butcher-Oemler effect is diminishing.  It is no longer apparent that
the star-formation rates seen in the distant BO galaxies are in any way
extreme given the characteristics of the surrounding field population
(Couch et al.\ 1997b).  While the only class of objects thought to be
uniquely associated with the BO effect in the distant clusters, the
post-starburst galaxy (PSG) or ``E+A'', have recently been discovered
in the local field (Zabludoff et al.\ 1996).  Hence, although our study
focuses on a sample of luminous X-ray clusters, our emphasis is to use
the X-ray properties of the clusters to provide a well-defined sample
of massive clusters and also to give some insight into their dynamical
states, rather than to simply test possible interactions between the
galaxy populations and the cluster X-ray gas. 

We briefly describe the structure of this paper.  Section 2 details the
observations and their reduction and analysis.  In section 3 we
catalogue the galaxy populations in our clusters before describing
our analysis of these catalogues to study the properties of the cluster
populations, both the evolved red spheroidal systems and the star
forming galaxies associated with the clusters. Finally, in section 4 we
discuss our results and give our main conclusions.

\section{Observations and Analysis}

The cluster sample we elected to use for this study is ideally suited
for investigating the properties of the galaxy populations in the most
massive collapsed structures at $z\sim 0.2$.  The sample contains a
total of 24 of the most luminous  X-ray clusters in the redshift range
0.2$<z<$0.3 selected from the northern section of the ROSAT All Sky
Survey, as given in an early compilation of the Brightest Cluster
Sample of Ebeling et al.\ (1996).  These clusters all have X-ray
luminosities in excess of $L_X($0.1--2.4$) \geq 2\times 10^{44} h^{-2}$
erg s$^{-1}$,\footnote{We take $h = H_\circ / 100$ kms/sec/Mpc and
$q_\circ = 0.5$ unless otherwise stated.} indicating that their typical
masses are: $M\geq M_{\rm Coma}$ and they thus represent some of the
most massive collapsed structures known.  This is also supported by the
richnesses of these systems as demonstrated by our analysis below.  The
X-ray luminosities derived from ROSAT HRI imaging are presented in
Table~1 and are discussed in more detail in Edge et al.\ (1997a, in
prep).  The sample has also been subjected to extensive X-ray
observations, providing deep, high resolution X-ray images of all the
clusters and X-ray spectroscopy of all but one.   These X-ray images
show that the majority of the clusters have relatively simple
morphologies, the most obvious exception being the bimodal cluster
A1758 (Fig.~1) which exhibits a highly elongated X-ray morphology.
Most importantly for the simple comparison of their galaxy populations,
the whole sample spans only a modest range in redshift, minimizing
differential K corrections.   We have obtained deep and  wide-field,
multi-colour imaging of a subsample of 10 of these clusters (those
whose right ascensions lie in the range 13--22$^{\rm h}$) for the
analysis here.  One particular feature of this project has been our use
of a blue sensitive CCD, allowing us to acquire deep $U$-band
(restframe $\sim 2900$\AA\ in the mid--UV) images of the clusters.  The
$U$ filter used for this study is a copy of Bessell's $U$ (Bessell
1990) kindly made available to us by Dr.\ J.\ Schombert. Our standard
Johnson $B$ and Cousins $I$ exposures provide complimentary information
on the cluster galaxies at restframe wavelengths around $\sim
3600$\AA\ and $\sim 6600$\AA, very close to $U$ and $R$ respectively.
  
The catalogue of candidate gravitationally lensed features from our
survey is given in Edge et al.\ (1997b), while the galaxy populations
in some of the individual clusters are briefly discussed in Smail et
al.\ (1995b) and Allen et al.\ (1997).  X-ray imaging of the clusters
in this sample are presented in Edge et al.\ (1997a).

\subsection{Observations and Reduction}

The data discussed here comprise $UBI$ imaging of all 10 clusters.
These data were acquired using the COSMIC imaging spectrograph
and a thinned 2048$^2$ TEK detector on the 5-m
Hale telescope at Palomar during the nights of June 9-12 1994 and July
5-7 1994.  This detector provides 0.284 arcsec sampling over a large,
$9.7 \times 9.7$ arcmin, field with very good response into the
near-ultraviolet.  The bulk of the data analysed here was taken in good
conditions during the June run.  The median seeing in the $I$-band
during these nights was 1.10 arcsec and the nights appeared stable and
photometric.  A log giving the field identifications, positions,
exposure times and seeing is presented in Table~1.  The reddening
values come from the estimated HI column densities in these fields,
using the conversion $E(B-V)=N(HI)/(4.8 \times 10^{21} {\rm cm}^{-2}$).  The
exposures typically comprise a 3000 s integration in $U$, along with
shorter 500 s integrations in $B$ and $I$.  These exposures were 
split into $\sim 2$--4 sub-exposures to allow for cosmic-ray rejection
during processing.  Observations of Landolt (1992) standard fields were
interspersed between the science exposures.

Standard IRAF reduction procedures were used to process both the
science frames and standards.  Firstly the images had their bias levels
calculated from the overscan region and this subtracted from the
exposure, which was then trimmed.  Flatfields were constructed from
either dome ($U$) or twilight observations ($B$ and $I$) to roughly
remove sensitivity gradients across the detector.  Having initially
flattened the frames, we then cleaned them of the brighter galaxies and
stars and used a median algorithm to stack all the cleaned frames in a
given filter for each night.  This created an illumination correction
frame which was smoothed with a large box filter, normalised and
divided through the science frames to remove any remaining mismatch
between the ``true'' sky illumination and that removed by the
flatfields.  These frames were then aligned and coadded with a
rejection algorithm to eliminate the cosmic rays.  Examples of the
final reduced images are given in Fig.~1.

%
%
\begin{figure*}
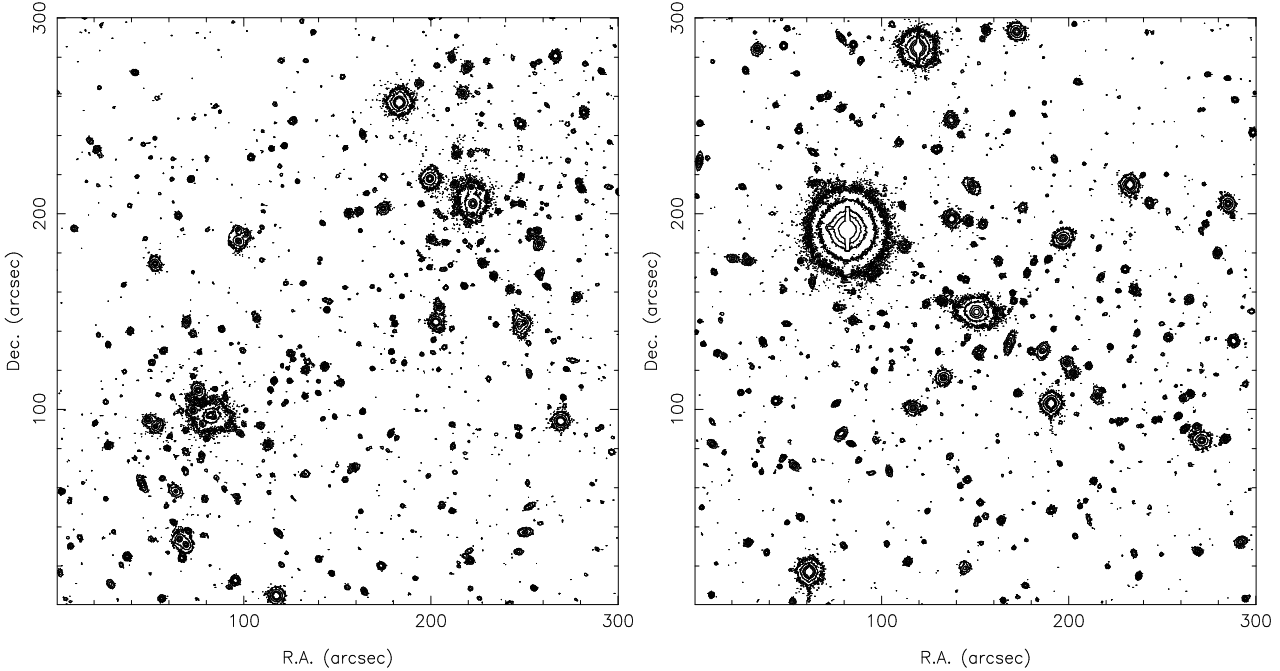

\begin{center}
\hbox{\psfig{figure=f1a.ps,height=3.5in,width=3.3in,angle=0}
\psfig{figure=f1b.ps,height=3.5in,width=3.3in,angle=0}}
\caption{$I$ band images of the central  $5\times 5$ arcmin regions of two of
the clusters in our sample, A1758 ($z=0.280$) and A1763 ($z=0.228$).
Note the two luminous galaxies at the upper-right and lower-left of the
field in A1758, these are the central galaxies of the two prominent
components of this bimodal cluster. The lowest contour corresponds to
$\mu_I=23.5$ mag arcsec$^{-2}$, equivalent to $3.5 \sigma$ pixel$^{-1}$.
}
\end{center}
\end{figure*}

Total exposure times and seeing from the final stacked images are
listed in Table~1.  Analysis of the standard star exposures through the
nights confirms our impression that all the nights used for this
project were photometric.   Photometric calibration of
our science exposures indicates average zero point errors contribute to
the scatter in the colours at the level of $\delta (U-B) = 0.052$,
$\delta (B-I) = 0.024$. Analysis of the colour terms for our
filter+detector combinations shows that these are negligible
compared to the reddening corrections (Table~1) and so we have applied
only the latter correction ($E(U-B)=0.85 E(B-V)$ and $E(B-I)=2.45
E(B-V)$).  Combining the uncertainty in the reddenings ($\pm 0.01$)
with the zero point errors, we estimate that our colours are accurate
to $\delta (U-B) = 0.053$ and $\delta (B-I) = 0.034$ on average.
We have tested the accuracy of our photometry (and the
reddening corrections) using the distribution of reddening-corrected
colours for objects identified as stars on our frames.  Comparing the distribution in \UB--\BI\ colours
for stars identified in each cluster to that given in Landolt (1992)
using a two dimensional K-S test we find only two instances of
significant ($>99$\%) offsets, these are in the $B$ photometry of A2390
($\Delta B = -0.14$) and A2261 ($\Delta B = 0.12$).  Applying these
corrections and then combining the stellar colour distributions for 314
stars in the ten clusters we determine limits on the maximum offsets in
our mean colours of $\Delta (U-B) = 0.02\pm 0.04$ and $\Delta (B-I) =
-0.04\pm 0.05$.  The errors on these offsets are consistent with those
claimed for the individual passbands.\footnote{This analysis
highlighted two stellar objects in our $I<20$ sample which had blue
\BI\ colours, but were very red in \UB.  These objects have $I=18.33$,
$(B-I)=2.52$, $(U-B)>4.7$ and $I=19.67$, $(B-I)=2.06$, $(U-B)>3.3$.
Their colours indicate that they are most likely to be $z\gs 3$ QSOs,
in which case we would estimate a space density of such objects of
$\sim 8$ degree$^{-2}$.  The two objects lie in A1758 ($\alpha({\rm
J2000})= 13~32~23.2$, $\delta({\rm J2000})= +50~34~32$) and A2261
($\alpha({\rm J2000})= 17~22~37.4$, $\delta({\rm J2000})= +32~11~19$), at
radii of 330 and 240 arcsec from the central galaxies. The
object in A1758 is also detected as an X-ray source in the PSPC/HRI
images of this cluster.}

Having aligned and calibrated our $UBI$ images we now create object
catalogues from the $I$-band images of each cluster.  For this we use
the  SExtractor package (Bertin \& Arnouts 1996) to detect and analyse
galaxies on the $I$ frames.  We adopt a detection criteria of 10
contiguous pixels each 1$\sigma$ (of the pixel-to-pixel noise) above
the local background, after convolution with a $3\times 3$ pixel
top-hat filter.  These catalogues are then visually inspected and
cleaned of spurious objects (e.g.\ diffraction spikes and noise objects
in the halos of a few very bright stars).  The coordinates from the
cleaned object catalogues so created are then used to measure colours
(\UB\ and \BI) for all the objects detected.  Before measuring colours,
however, we first match the seeing in the $UBI$ frames from the
profiles of stars in the images.  We have chosen to use a fixed angular
size photometry aperture, 3.0 arcsec in diameter (6.8--7.9 $h^{-1}$ kpc),
rather than a fixed metric aperture to simplify the field corrections
for our catalogues.  We discuss the uncertainties which arise from this
later in the paper.  These apertures are sufficiently large that we
should detect the bulk of the light from both bulge and disk components
of luminous galaxies in these clusters.  The final frames typically
cover an area of $\sim 90$ sq.\ arcmin to an 80\% completeness limit of
$I=22.5$--23.0 (Table~1), where these values are estimated from
comparing the observed galaxy counts with those expected from deeper
field observations (Smail et al.\ 1995a).  To determine galaxy
luminosities we use the {\sc BEST} magnitude provided by SExtractor,
which is based on a ``Kron-type'' magnitude for brighter galaxies, with
a fixed minimum aperture size for fainter objects.  In our analysis we
restrict ourselves to an $I\leq 22$ sample to ensure $\gs 95$\%
completeness in all our fields.  Analysis of the other passbands shows
that our exposures provide photometry to 20\% accuracy to median limits
of $U=25.1$ and $B=25.9$.  The final catalogues contain a total of
11,211 objects brighter than $I=22.0$ over an area of 898 sq.\ arcmin,
of which 1138 have profiles consistent with being stars, giving an
average surface density of $11.2\pm 2.3$ galaxies arcmin$^{-2}$, where
the scatter is field-to-field.
  
\subsection{Colour Distribution in the Field}

To determine the field correction for our cluster frames we used the
deep $U$ exposure analysed by Hogg et al.\ (1997).  This 28.0 ksec
exposure covers an area of 81.0 sq.\ arcmin in a high-latitude blank
field (Table~1) and was taken with the same instrument as used for our
observations, although using a different $U$ filter.  To supplement
this observation we acquired 500s exposures of the field in $B$ and $I$
during the night of January 31 1997 (Table~1).  These frames were
reduced in the same manner as the cluster observations.  Calibration of
these data was provided from observations of Landolt (1992) standard
stars, observed before and after the science exposures, giving zero
point errors of $\Delta B= 0.04$ and $\Delta I= 0.05$.  We confirm the
accuracy of the colours determined from these fields using the locus of
stars on the colour-colour plane as was done for the cluster
observations. 

The objects in this field were catalogued in a similar manner to the
cluster observations, starting with an $I$-selected sample and
measuring colours in seeing-matched apertures.  The precision of these
colours is similar to that achieved in our cluster observations due to
equivalent exposure times for the $B$ and $I$ observations.  The final
catalogue contains 377 galaxies brighter than $I=22.0$, or a surface
density of $4.7\pm 0.5$ galaxies arcmin$^{-2}$, where the variation is
estimated from independent sub-regions within the field.  As we will
show the clusters studied here are sufficiently rich, given their
redshifts, that field contamination only becomes an important
correction at the faintest magnitudes discussed.  To $I=20$ the typical
cluster frame contains 315 galaxies, of which 90 (28\%) are expected to
be field contamination, predominantly blue galaxies, this fraction
increases to 40\% at $I=22$, the faintest limit used in this work.  In
our analysis we will ignore the effects on the estimated field
correction of gravitational amplification of the background field
population.  The majority of our analysis focuses on galaxies with red
optical colours and for these we expect the effects of lensing to
actually lower the background field density (c.f.\ Broadhurst et
al.\ 1997), leading to a slight over-correction for field
contamination.

\section{Analysis and Results}

We start by describing the general colour distribution of 
galaxies within our clusters, before identifying particular
groups of objects and discussing their properties in more
detail in the following sections.  

%
%
\begin{figure}
\begin{center}
\psfig{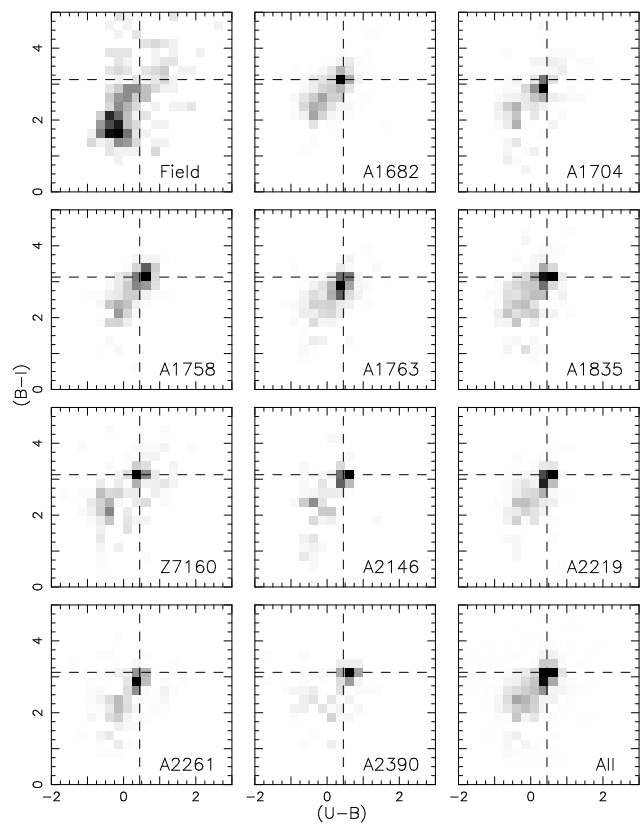}
\caption{The \UB--\BI\ colour-colour plane for each cluster showing the
distribution of colours for galaxies brighter than $I=22.0$, after
correction for field contamination.  We also show the equivalent
distribution for the field population, as well as the combined
distribution for all the clusters.  The dashed lines are to facilitate
the comparison of the various distributions.
}
\end{center}
\end{figure}

We show in Fig.~2 the distribution of galaxies on the \UB--\BI\
colour-colour (c-c) plane in each of the 10 clusters, after correcting
for field contamination.  We also show the composite colour
distribution for all the clusters combined.  All of these plots use
only those objects classified as galaxies and with magnitudes brighter
than $I\leq 22.0$.  No differential K corrections, to compensate for
the different cluster redshifts, have been applied to any of these
plots.  In Fig.~3 we present the combined cluster sample as three
independent magnitude slices, along with the whole sample.  On these
plots we have marked the locus of colours expected for the spectral
energy distributions (SED) representative of the spectral types of
galaxies with E, Sab, Sbc, Scd and Sdm morphologies in the local
Universe (effectively a series of increasing ratio of current to
average star-formation rate), as they would be observed at $z=0.24$.
It can be seen that the clusters contain galaxies spanning the whole
range of star-formation rates observed in the local field.  At bright
magnitudes, however, the distribution is dominated by a population of
galaxies whose \UB\ and \BI\ colours similar to unevolved
elliptical galaxies at these redshifts.  As we look at fainter
luminosities within the clusters we find a gradual increase in the
number of blue galaxies, with colours ranging across those expected
from galaxies with star-formation rates similar to local Sbc to Sdm
galaxies.  A similarly wide range in galaxy colours is also seen for
the brightest cluster galaxies (Fig.~3), these are typically D or cD
galaxies, and although a number of them show the red colours expected
for such massive spheroidal systems, three show substantially bluer
colours indicating significant on-going star-formation (Allen 1995).

%
%
\begin{figure*}
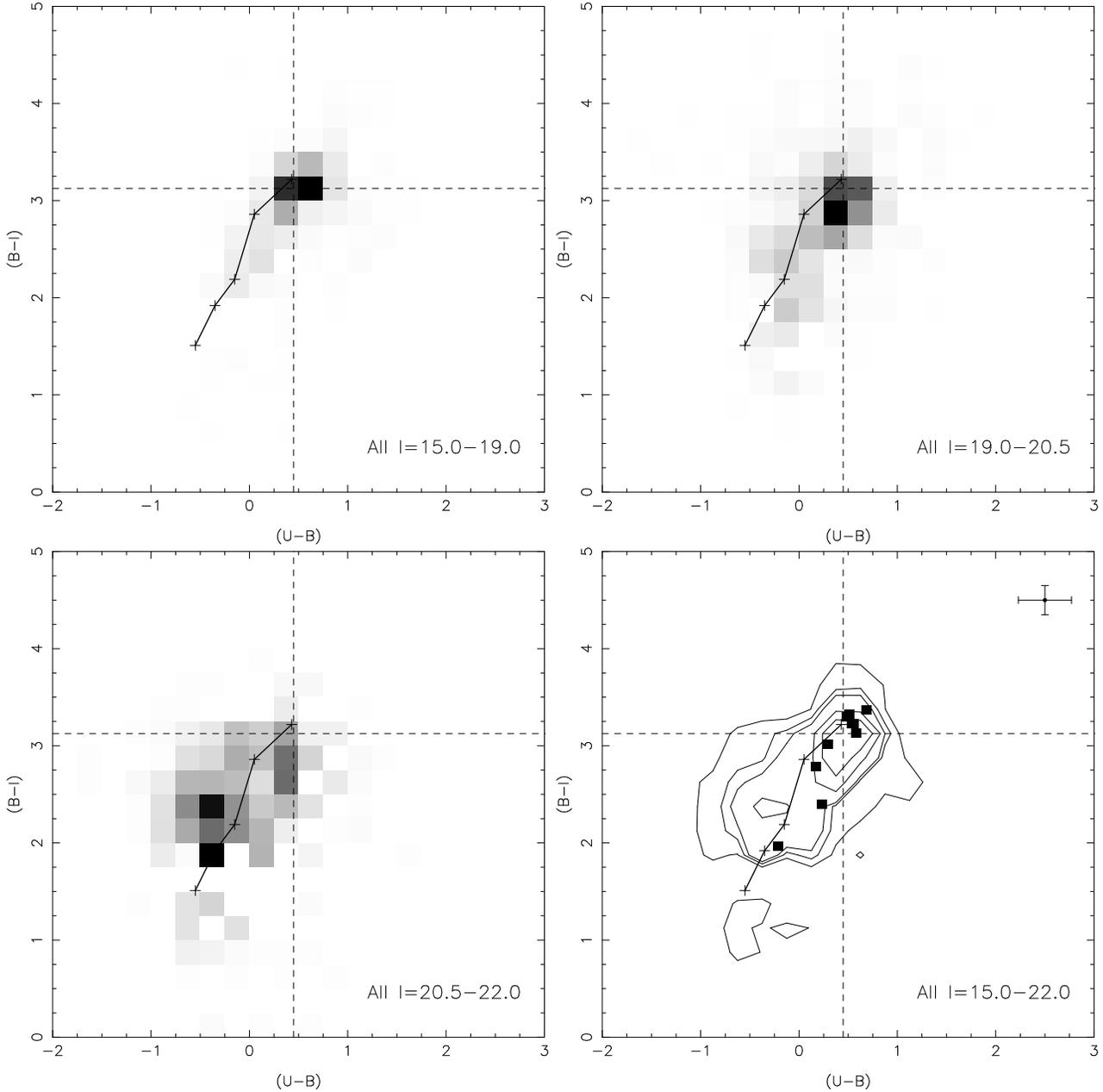

\begin{center}
\hbox{\psfig{figure=f3a.ps,height=3.3in,width=3.3in,angle=0}
\psfig{figure=f3b.ps,height=3.3in,width=3.3in,angle=0}}
\hbox{\psfig{figure=f3c.ps,height=3.3in,width=3.3in,angle=0}
\psfig{figure=f3d.ps,height=3.3in,width=3.3in,angle=0}}
\caption{The combined \UB--\BI\ distribution for all 10 clusters after field
correction.  We show this for three independent magnitude slices,
$I=15.0$--$19.0$, $I=19.0$--$20.5$ and $I=20.5$--$22.0$, as well as for
the whole sample with $I=15.0$--$22.0$.  We also over-plot the locus of
colours for the non-evolved spectral energy distributions corresponding
to local spectral types E, Sab, Sbc, Scd and Sdm (upper-right to
lower-left).  On the $I=15.0$--$22.0$ panel the point at the upper left
illustrates the 80\% upper limits on the photometric errors at $I\leq
22$.  In addition on this panel we show the colours for the central
galaxies of the clusters, measured within a $\sim 7.5 h^{-1}$ kpc
diameter aperture.  The three central galaxies with anomalously blue
colours are (from the bluest in \BI) in:  A1835, A2390 and Zw7160.  All
these central galaxies also show line emission indicating some level of
on-going star-formation (Allen 1995).  The contours in the final plot
are [0.1,0.5,0.8,2,4,8,12]~$\times 10^3$ galaxies mags$^{-2}$.  
}
\end{center}
\end{figure*}

We now identify three groups of cluster galaxies on the
\UB--\BI\ ~colour-colour plane to $I=22$ and use these to highlight the
various components of the cluster populations. At this depth the median
photometric errors are $\delta(U-B)=0.11$ and $\delta(B-I)=0.06$, with
80\% of the objects having errors less than $\delta(U-B)=0.27$ and
$\delta(B-I)=0.15$, adequate to define independent regions of the
colour-colour plane.   In particular the accuracy of our \BI\
photometry allows us to robustly separate samples of galaxies using
this colour, which measures the strength  of their 4000\AA\ breaks.
The three groups we identify are: 1) a large population of galaxies
with colours similar to passive elliptical galaxies at the cluster
redshift; 2) the cluster population with colours typical of star-forming galaxies at $z\sim 0.24$; 3) a number of galaxies whose
\BI\ colours indicate a strong 4000\AA\ break, representative of an
evolved stellar population, but which have bluer \UB\ colours than
group \#1 showing that they are, or recently have been, forming stars.
For ease of use we have called these three groups:  ``Red'', ``Blue''
and ``UV+''.   To estimate the relative proportions of these groups
within the clusters we define rough limits of their boundaries on the
\UB,\BI colour-colour plane as: [0.1:0.9,2.9:3.6] for Red,
[$-$1.1:0.9,1.9:2.9] for Blue and [$-$1.1:0.1,2.9:3.6] for UV+.  Note
that the exact definitions of the samples used in our analysis below
vary according to what questions we are attempting to answer, we have
used the regions above simply to illustrate the distribution of objects
in the various groups.  The typical proportions of the three cluster
populations are then: [0.95,0.05,0.00] to $M_V= -18.5 + 5 \log h$ and
[0.66,0.29,0.05] to $M_V= -17.0 + 5 \log h$.  We give the proportions
in the individual clusters in Table~2.  The average radial
distributions of the three galaxy populations in the clusters within
500 $h^{-1}$ kpc of the cluster centres are shown in Fig.~4.  Outside
of $\sim 100 h^{-1}$ kpc the profile of the Red galaxies drops as
$\alpha = -0.96\pm 0.08$, close to the value expected for an isothermal
distribution.  The UV+ population follows a
radial distribution close to that of the red galaxies, although with a
large scatter, $\alpha = -0.87\pm 0.46$.
While the Blue galaxy distribution is considerably
flatter with $\alpha = -0.46\pm 0.20$.  

Finally, we can identify one further sample of galaxies on the c-c
plane:  those galaxies with colours substantially redder than the
cluster elliptical sequence in \BI\ (which provides a good measure of
the strength of the 4000\AA\ break at the cluster redshifts).  This
last group consists of galaxies which are probably background to the
cluster, as the elliptical sequence represents galaxies with the
strongest 4000\AA\ breaks at the cluster redshift, and as such we do
not discuss them further here.  The number of objects lying substantially
redward of the cluster sequence in \BI\ is consistent with the 
number expected from our field distributions. 

%
%
\begin{figure}
\begin{center}
\psfig{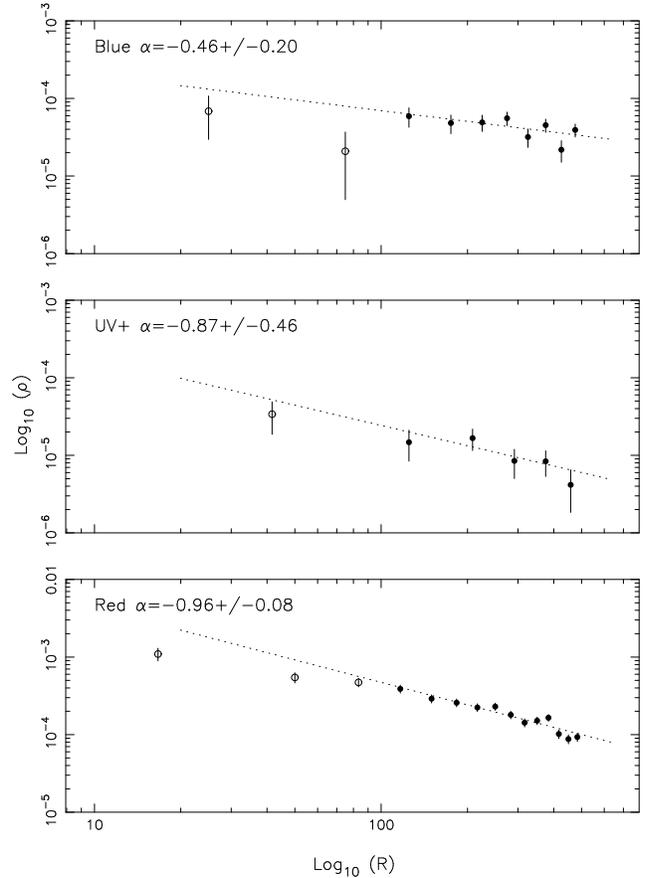}
\caption{The radial density distributions (in
numbers of galaxies per sq.\ $h^{-1}$ kpc) of the three galaxy
populations (``Red'', ``Blue'' and ``UV+'') identified off the colour-colour
planes in Fig.~3.  These are shown for a magnitude limit of $M_V= -17.0
+ 5 \log h$ and have been corrected for field contamination.  The 
bimodal cluster, A1758, is not used
in this plot.  We fit power laws to the radial distributions
outside of 100 $h^{-1}$ kpc, to reduce the effects of crowding on the fits,
the points used are marked as filled symbols. 
The best fit power laws are plotted on the figures.  
}
\end{center}
\end{figure}

\subsection{Red Cluster Galaxies}

The most obvious feature of the \UB--\BI\ plots in Figs.~2 and 3 is the
strong clump of red galaxies in all the clusters.  The clump is
populated with the luminous elliptical galaxies which dominate the
bright end of the cluster population.  Replotting the catalogues as
colour-magnitude (c-m) diagrams illustrates another property of this
population -- a very narrow colour-luminosity relation in both \UB--$I$
and \BI--$I$ (Fig.~5).  In both panels the elliptical sequence slants
down, indicating that the spheroidal galaxies have bluer colours at
fainter luminosities, a result of lower mean metallicities in the low
luminosity galaxies (Kodama \& Arimoto 1997).  In the fainter samples
in Fig~3, however, this population appears to disperse, a feature which
may be associated with the apparent ``break'' at $I\sim 20.5$ in the
red elliptical sequence on the \UB--$I$ plane. We return to discuss
this feature later.

%
%
\begin{figure*}
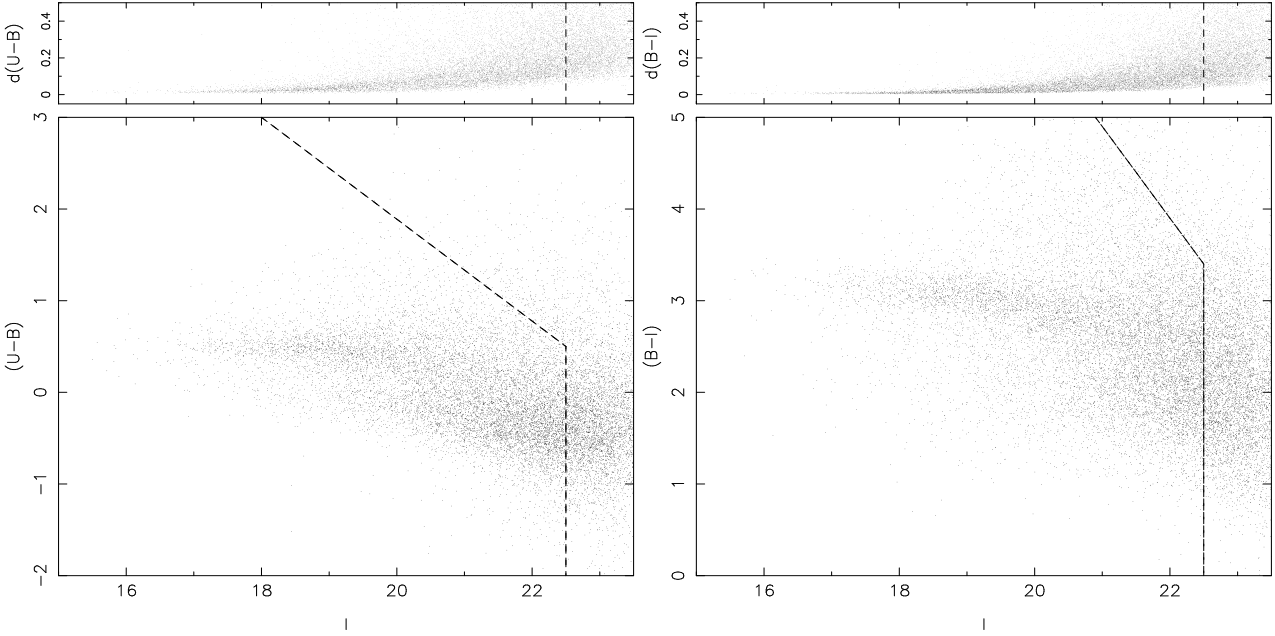

\begin{center}
\hbox{\psfig{figure=f5a.ps,height=3.3in,width=3.3in,angle=0}
\psfig{figure=f5b.ps,height=3.3in,width=3.3in,angle=0}}
\caption{The colour-magnitude distributions for the combined sample of
galaxies from the 10 clusters.  We have made no attempt to correct
the colour-magnitude distributions from the various clusters for
their different redshifts and so some blurring of features is
expected in this plot.  The strong linear feature visible at
bright magnitudes in both
plots is due to the red elliptical cluster members.  Notice the apparent
decline in the number of galaxies lying on this sequence in the \UB--$I$ 
plot at $I\gs 20.5$.  The dashed lines show the 80\% completeness limits
for the various catalogues, these roughly correspond to a typical
photometric accuracy of $\sim 0.2$ mag in each band.
No correction has been made for field contamination in this figure.}
\end{center}
\end{figure*}

The first question we wish to investigate is the variation in the
typical colours of the brighter spheroidal galaxies across the
clusters. As we discussed in the introduction, the dispersion in the
restframe UV--optical colours of galaxies is a sensitive test of their
recent star-formation, and the variation of the mean colours for the
spheroidal population can give powerful constraints on their formation
epoch  (BLE; Ellis et al.\ 1997) as well as the formation of the larger
structures they inhabit.  To study this scatter we fit to the
colour-luminosity relation in our clusters and then correct these fits
to a fiducial cluster at $z=0.24$ (the mean of our sample) to compare
the various values.  

We fit a linear relation ($(U-B) =A_{(U-B)}+B_{(U-B)} I$ or the
equivalent for \BI) to the c-m relation using the same technique
applied by BLE and Ellis et al.\ (1997):  Huber's robust estimator,
after removal of probable field contamination.  Galaxies are rejected
from the fit interactively until the fit is stable to the inclusion of
remaining points, the expected number of field galaxies is used as a
guide to indicate the numbers of galaxies to be removed (typically
$25\pm 5$ galaxies, or $\sim 10$\% of the sample). We have confirmed
that these fits are not sensitive to modest changes in the definition
of the red sub-sample region on the c-c plane.  We then fit the
resulting sample across a fixed range in absolute luminosity
corresponding to $M_V - 5 \log h = [-23.5,-18.5]$ ($I\ls 20$).  For
simplicity we parameterise the variation in apparent magnitude across
the redshift range of our sample, for a $M_V = -18.5 + 5 \log h$ galaxy
with a non-evolved elliptical SED the apparent magnitude is: $I=17.30 +
10.1 z$.  The fits are evaluated at an absolute magnitude corresponding
to $I=18$ at $z=0.24$, roughly $M^\ast$.  We then apply differential K
corrections to the intercept values to convert them to a cluster
observed in \UB\ or \BI\ at $z=0.24$, the corrections assume a
non-evolved elliptical SED and are parameterised as:  K$(U-B) = -1.11
\Delta z$ and K$(B-I) = -3.14 \Delta z$ to 2\% accuracy.  These
corrections amount to $\ls 0.05$ in \UB\ and $\ls 0.13$ in \BI, with
uncertainties of $\ls 0.003$ at a fixed luminosity.  The different K
corrections associated with the variation in galaxy colour along the
elliptical sequence will introduce a scatter into the corrected
colours.  ~From the slope of the c-m relations we expect this to be at
the level of $\delta B_{(U-B)} \sim 0.008$ and $\delta B_{(B-I)} \sim
0.006$, where the slope is steeper in both colours at higher
redshifts.  This is substantially larger than the colour variations
introduced between clusters due to our adoption of a fixed aperture
size for our photometry, $\delta \ls 0.001$, and so we ignore the
effects of the variation in metric aperture size across the different
clusters.  We obtain the fits listed in Table~3a and 3b (for \UB\ and
\BI\ respectively). ~From these we determine average slopes for the c-m
sequence of $B_{(U-B)} = -0.041\pm 0.020$ and $B_{(B-I)} = -0.067\pm
0.019$, where the scatter within the sample is comparable to the
typical errors on the fit.  We have therefore simply adopted these
average slopes for all the clusters.

%
%
\begin{figure}
\begin{center}
\psfig{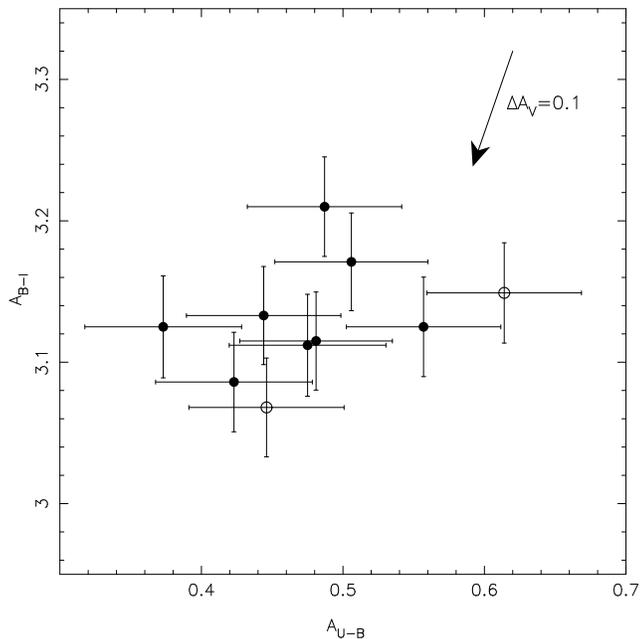}
\caption{The intercepts for the c-m relations in the different
clusters (Table~4), evaluated at a fiducial magnitude equivalent to $I=18$ at
$z=0.24$ and corrected for differential K corrections assuming a
non-evolved elliptical SED.  The two open symbols mark A2390
and A2261, where the photometry was corrected using the stellar
sequences in the CCD frames. The vector indicates the effect of changing the
adopted reddening and we have included the expected random calibrations
errors for the clusters in their error bars.}
\end{center}
\end{figure}

Next to simplify the comparison of the clusters we fix the relevant
slopes at these average values and refit to our data, across two
absolute magnitude ranges:  $M_V - 5 \log h = [-23.5,-18.5]$ ($I\ls
20$) and a more restrictive range $M_V - 5 \log h = [-23.5,-20.0]$
($I\ls 18.5$).  The results of these fits (evaluated at $I=18$) are
listed in Table~4 and we illustrate the values for the complete sample
in Fig.~6.   These values show a remarkably small dispersion, $A_{(U-B)} =
0.481\pm 0.068$ and $A_{(B-I)} = 3.129\pm 0.041$, while for the
smaller, bright subsamples we obtain $A_{(U-B)} = 0.487\pm 0.087$ and
$A_{(B-I)} = 3.130\pm 0.046$.  We find no statistically significant
correlation between the mean colours in the two passbands. 

The typical colours of the bright elliptical cluster population derived
above are similar to those expected from a local L$^\ast$ elliptical
galaxy placed at $z=0.24$: $(B-I)=3.17$ and $(U-B)=0.42$.  The observed
colours are offset by $\Delta_{(U-B)} = 0.05\pm 0.07$ and
$\Delta_{(B-I)} = -0.05\pm 0.04$ compared to these.  These offsets are
in a similar sense to the limits we placed on the relative offset of
the colours of stars in our fields compared to the distribution in
Landolt (1992).  Taking the uncertainty in our absolute colour system
into consideration we can only place weak limits on the extent of
colour evolution of cluster ellipticals out to $z=0.24$. Looking at the
average \BI\ c-m slope we derived from our clusters, we can crudely
compare this with that measured in $(U-R)$ using the slope observed in
$(U-V)$ for spheroidal galaxies in Coma by BLE.  They find
$B_{(U-V)}=-0.082\pm 0.008$, the expected slope in $(V-R)$ is
$B_{(V-R)}\sim -0.020$ (Kodama \& Arimoto 1997), indicating a predicted
slope of $B_{(U-R)}=-0.10\pm 0.01$, only slightly steeper than the
observed value in our distant clusters of $B_{(B-I)}=-0.07\pm 0.02$.
Unfortunately, there is little information on the UV properties of
cluster ellipticals shortward of the atmospheric cut-off and so we
cannot compare the observed slope in \UB\ with a locally determined
value.

Turning to the scatter in the mean colours of the elliptical
sequences between the clusters.  The random errors in the calibration
of the clusters are approximately $\delta(U-B)= 0.053$ and
$\delta(B-I)= 0.034$, removing these contributions from the observed
scatter would indicate intrinsic $1\sigma$ dispersions of $\delta (U-B)
\ls 0.04$ and $\delta (B-I) \ls 0.02$. We therefore conclude that the
red, spheroidal populations brighter than $M_V=-18.5 + 5 \log h$ are
remarkably homogeneous across all 10 clusters in our sample at a
lookback time of $\sim 2 h^{-1}$ Gyrs.    We also note that
the small spread in colours for the brighter ellipticals in the
clusters would allow them to be used as a crude, but economical,
redshift indicator provided the colour evolution of the population can
be calibrated for a number of clusters spanning the redshift range of
interest.  The observed scatter in the mean \BI\ colour of the bright
cluster ellipticals, including measurement errors, is $\sim 0.04$,
corresponding to an accuracy in the estimated redshift of $\Delta z
\sim 0.01$ (c.f.\ Belloni et al.\ 1995). 

A complete analysis of the expected scatter in the mean colour of the
elliptical sequences within rich clusters will require a combined
spectral modelling and Press-Schechter approach (e.g.\ Baugh et
al.\ 1996). Nevertheless, we can use a simple, if somewhat contrived,
model to understand what general constraints our observations provide
on the discreteness of the structures which coalesce to form rich
clusters.  In the model, clusters form from sub-units (groups) of
galaxies each containing a fixed number of elliptical galaxies, $N_E$.
The elliptical galaxies within a given groups have synchronised
formation at some epoch (this could be thought of as relating to the
collapse redshift of the group) which we distribute uniformly in time
between the Big Bang and a cut-off time, $t_{\rm End}$, when elliptical
formation ceases in all structures.  We define $t_{\rm End}$ relative
to $z=0.24$ and hence $t_{\rm End}=0$ means that elliptical formation
continues right up to $z=0.24$, while $t_{\rm End}=5$ Gyrs ($h=0.5$) corresponds
to a time 5 Gyrs prior to $z=0.24$ or roughly $z_F\sim 1$.  In the
Press-Schechter models the redshift cut-off and the distributions in
relative formation times for ellipticals within the groups would
probably both be functions of group mass, as well as cosmology, but the
current model is sufficient to illustrate the technique.  We now
determine what joint constraints our observed cluster-to-cluster
scatter in \BI, restframe $(U-R)$, provides on the numbers of
structures from which the clusters were built, and the most recent
epoch of elliptical formation in these structures.  We use the spectral
evolution model of Fioc \& Rocca-Volmerange (1997) for elliptical
galaxies to determine the evolution in the rest-frame $(U-R)$ colours
of the galaxies.  Starting from the observed number of ellipticals in
the cluster, we randomly select groups of $N_E$ galaxies and assign
them a formation epoch uniformly in the available time interval, from
this we determine their $(U-R)$ colour at $z=0.24$.  We repeat this
until we have fully populated the cluster and then calculate the mean
colour of the elliptical population in the cluster before moving onto
the next cluster,  until all have their full complements of elliptical
galaxies. With values for the mean colours of the elliptical population
in each of the 10 model clusters we next estimate the scatter in the
colours {\it between} the clusters.  This proceedure is repeated 10,000
times for each combination of $N_E$ and $t_{\rm End}$ to estimate the
likelihood that the scatter amongst the 10 model clusters would be less
than or equal to that observed for those parameters.

%
%
\begin{figure}
\begin{center}
\psfig{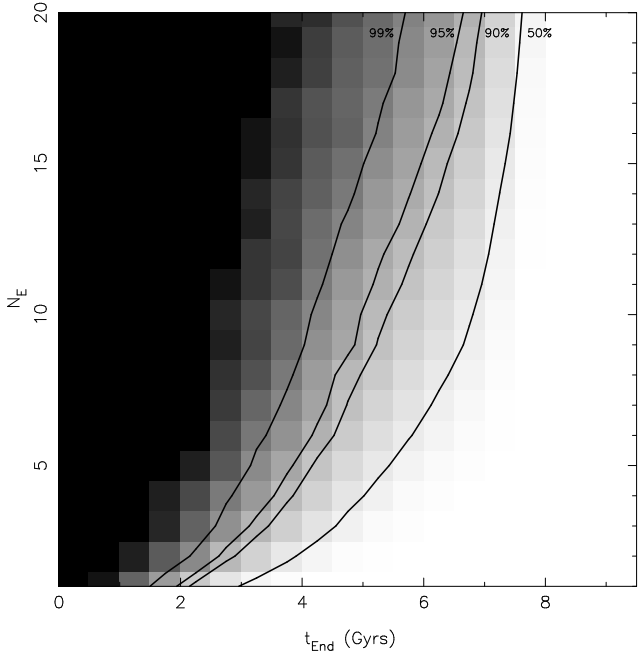} 
\caption{The logarithmic likelihood distribution for the two free
parameters of our simple model of cluster formation: $N_E$, the number of
ellipticals in a typical group accreted by the cluster; and $t_{\rm
End}$, the time before $z=0.24$ when elliptical formation stops
in all structures.  The contours show the probabilities that the
scatter between the clusters would be less than the observed
value, and mark the 50, 90, 95 and 99\% confidence limits.  Thus the
lighter area to the right of the figure indicates the allowed region of
the parameter space.  For example if the clusters form from groups
containing $\sim 10$ co-eval ellipticals, $N_E=10$, then these
structures must have formed their galaxies earlier than $t_{\rm End}\gs
4$--5 Gyrs before $z=0.24$ (95\% c.l.), equivalent to $z_F \gs 1$.  Here
we have used the brighter sample of cluster ellipticals (Table~4) to
constrain the model and assumed a $h=0.5$ and $\Omega=1$ cosmology
which gives an age of the universe of 9.5 Gyrs at $z=0.24$.}
\end{center}
\end{figure}

Looking at Fig~7 it is apparent that as we form the cluster from larger
and larger sub-units (higher $N_E$) then the scatter between the
clusters should increase.  Remembering that all the galaxies within a
given sub-unit form together, but that the sub-units themselves can
form at any point within the available time period, we can see that
this constraint arises from the increasing shot-noise in the average
colour of the cluster population when it is formed from a smaller and
smaller number of independent sub-units.  At one extreme where each
cluster forms as a single group of ellipticals, $N_E\gg 1$, which
themselves all formed together at a given time, then the scatter
between the clusters simply reflects the scatter between the formation
times of the original groups and these are therefore constrained to all
have formed in a relatively short period of time (to minimise the
scatter between their present day colours).  Such a model might
represent the result of a top-down formation of clusters from large
Zel'dovich pancakes with the bright ellipticals being formed in the
first collapse. Looking at Fig.~7 we could constrain the elliptical
formation to occur at least $t_{\rm End} \sim 6$--7 Gyrs ($h=0.5$)
before $z=0.24$ or $z_F \gs 2$.  At the other extreme, where the
clusters form from accreting individual elliptical galaxies, $N_E\sim
1$,whose formation epochs are uncorrelated, we see that given the large
number of ellipticals within each cluster the mean colours of the
clusters have a very small scatter and hence the constraint on the last
period of star-formation in the elliptical population is much weaker
(equivalent to  $z_F \gs 0.5$).  Taking an intermediate case, if the
clusters typically form from sub-clumps and groups containing $\sim 10$
ellipticals, and the formation of the galaxies within each group is
synchronised, then these structures must have formed their galaxies at
least 5 Gyrs ($h=0.5$) before $z=0.24$ (95\% c.l.), equivalent to
$z_F\gs 1$.  Requiring that the bulk of the ellipticals form much
earlier than this (e.g.\ $z\gs 3$, Ellis et al.\ 1997) would mean that
we could only rule out their coherent, stochastic formation in large
structures ($N_E\sim 20$).

The colours of the bright cluster ellipticals indicate that this
population is remarkably homogeneous between clusters. We can further
test this homogeneity by investigating the scatter in the amount of
baryonic material locked up as stars in the cluster ellipticals, as
compared to the total mass of the cluster.  The simplest method to
achieve this is to compare the luminosity in the elliptical sequence to
the cluster X-ray temperatures for the 7 clusters with published
temperatures (Mushotzky \& Scharf 1997).  We determine the total
spheroidal luminosity in our cluster by simply integrating the light in
the cluster sequences defined above down to $M_V=-18.5 + 5 \log h$
across the whole field (an effective radius of $0.75 h^{-1}$ Mpc in a
typical cluster). The values obtained, along with a bootstrap estimate
of their uncertainties, are given in Table~5.  We find a reasonable
linear correlation (Fig.~8) between the  the integrated red galaxy
luminosities and the X-ray temperatures: $L_{E} = (0.47\pm 0.08) \times
10^{12} T_X$, where the fit has been constrained to pass through
[0,0].   The dispersion around this line is small, only 17\%, (compared
to $\sim 30$\% when $L_X$ is used in place of $T_X$) indicating that
the integrated luminosity of the red galaxy population is a good tracer
of the total mass of the cluster.   In fact the ``true'' scatter may be
less when proper account is taken of the roles of mergers and cooling
flows in changing the observed X-ray temperatures.  Parameterising the
ratio of the cluster mass to the luminosity of the elliptical
population as: $M/L_E = \gamma h (M/L_V)_\odot$, and assuming an
isothermal gas distribution with $\beta_{\rm fit}=1$ we find $L_{E} =
(81 \times 10^{12} T_X) / \gamma $ for $L_E$ in solar units and $T_X$
in keV. Thus the observed best fit slope corresponds to $M/L_E =
(170\pm 30) h$ in the restframe $V$-band.  Including the observed
passive evolution of the stellar populations (see \S3.2) this
corresponds to an equivalent $z=0$ value of $M/L_E = (220\pm 40) h
(M/L_V)_\odot$.  This compares well to the value for the Coma cluster
of $M/L_V = 240 h (M/L_V)_\odot$ within $0.5 h^{-1}$ Mpc (transformed
from the $B$-band value of Fusco-Femiano \& Hughes 1994).

%
%
\begin{figure}
\begin{center}
\psfig{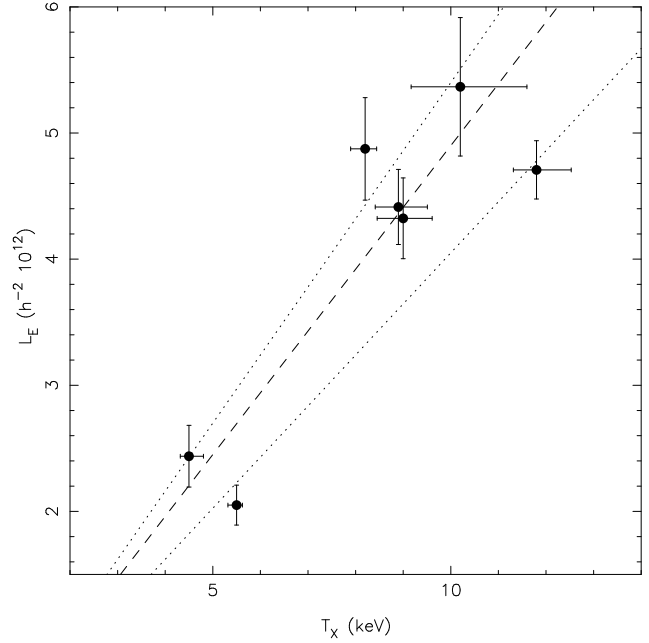}
\caption{The correlation between the integrated optical luminosities of 
the elliptical
populations in 7 of the clusters and their X-ray temperatures.   Errors are $1\sigma$ for both observables.
The dashed line shows the best fit linear
relationship (constrained to pass through [0,0]).  The slope of
this line corresponds to a $V$-band restframe $M/L_E = (170\pm 30) h$ 
in solar units within an
effective radius of $0.75 h^{-1}$ Mpc, the upper and lower dotted
lines indicate the relationships for $M/L_E = 150 h (M/L_V)_\odot$ and $M/L_E = 200 h (M/L_V)_\odot$ respectively.}
\end{center}
\end{figure}

\subsection{The ``UV+'' Cluster Population}

While the proportion of galaxies falling in the UV+ region of the
Fig.~3  is small at bright magnitudes, constituting only $\sim 0.5$\%
of the cluster population brighter than $M_V\sim -18.5 + 5\log h$, it
substantially increases as we reach fainter into the cluster population
(Table~2).  The colour selection used to define this sample is
sufficiently wide that it will include all cluster galaxies with strong
4000\AA\ breaks (E--S0--Sa), which also show blue \UB\ colours.   The
accuracy of our \BI\ colours means that we expect little contamination
of the sample from later spectral types (with bluer
4000\AA\ colours).  We now investigate the relation between this
population and the apparent break in the  c-m relation for the cluster
ellipticals in \UB--$I$.  To do this we first look at the luminosity
distribution of galaxies lying along the elliptical c-m sequences
defined above.  We select a wedge-shaped region parallel to the c-m
sequence with a width varying from $\Delta_{(B-I)}=0.18$--0.33
($\Delta_{(U-B)}=0.28$--0.43) from $I=16.0$--22.0, the increasing width
of the region at fainter limits compensates for the increased
photometric errors.  These regions contain 90\% of the morphologically
classified E and S0 galaxies brighter than $I=20.5$ in the HST/WFPC-2
image of A2390 (see \S 4). Using these areas we determine counts from
the combined cluster sample and correct these using the same regions of
the field galaxy colour-magnitude plane.  The magnitudes and colours
for the cluster galaxies are corrected to those of a fiducial cluster
at $z=0.24$ using the K corrections suitable for a non-evolved
elliptical SED (consistent with the typical colours observed for this
population).  

We plot the luminosity distributions determined from both
colour-magnitude planes (Fig.~9).  To parameterise the differences
between these two LFs we have fitted composite Gaussian+Schechter
functions (e.g.\ Wilson et al.\ 1997) to the distributions and these
are over-plotted in  Fig.~9.  The only free parameter in the fit is the
faint end slope of the Schechter function.  The parameters of the
Gaussian  function ($I_{cent} \sim 19.1$ and $\sigma \sim 1.4$) have
been fixed to agree with the luminosity function of elliptical galaxies
in local clusters (Binggeli, Sandage \& Tammann 1985, BST; Thompson \& Gregory 1993; Biviano et
al.\ 1995), after allowance is made for the expected luminosity
evolution of this population.\footnote{To determine the amount of
luminosity evolution in the bright galaxies clusters we use a simple
Schechter function fit to the luminosity distribution and then compare
the value of $M^\ast$ we derive with that given by Colless (1989) for
the distributions in 14 rich clusters, where he estimated $M^\ast_{b_j}
= (-19.84\pm0.06) + 5 \log h$ for $\alpha=-1.25$, or $M^\ast_V \sim
-20.5 + 5 \log h$ assuming a mean color of $(b_j-V)\sim0.7$.  We
therefore fit a single Schecter function to our distributions having
fix  $\alpha=-1.25$ and estimate $I^\ast =17.40\pm 0.07$ from our
\UB\ distribution and $I^\ast=17.42\pm 0.06$ from \BI.  These
correspond to $M_V = (-20.8\pm 0.1) + 5 \log h$, or $\sim 0.3\pm 0.1$ mag
brighter than the $z= 0$ value.  This is in good agreement with the
expectation from luminosity evolution of a passively evolving stellar
population formed at high redshift ($\Delta M_V \sim -0.3$, \S 3.3;
Barger et al.\ 1997).} The characteristic magnitude of the Schechter
function is fixed at $I=20.5$, $M_V = -17.7 + 5 \log h$. The main
feature of interest in Fig.~9 is the difference in the relative numbers
of  galaxies lying on the red cluster sequence beyond $I\geq 20.5$,
there being considerably fewer in the \UB\ sequence.  This is
quantified by the different values of the faint end slope, $\alpha$,
determined for the two samples, $\alpha_{(U-B)} = -0.40\pm 0.13$ and
$\alpha_{(B-I)} = -0.97\pm 0.07$.  Thus we confirm the visual
impression from the \UB--$I$ c-m diagram that there is a decline in the
number of red sequence members at magnitudes fainter than $I\sim 20.5$
($M_V\sim M^\ast + 2.5$).  Splitting the \UB--selected sample on the
basis of radius from the cluster center we see at most a very marginal
preference for the decrement to be greater in the cluster centre
($r\leq 250 h^{-1}$ kpc).  This is as expected given the similarity of
the radial profiles of the UV+ and Red galaxies shown in Fig.~4.

%
%
\begin{figure}
\begin{center}
\psfig{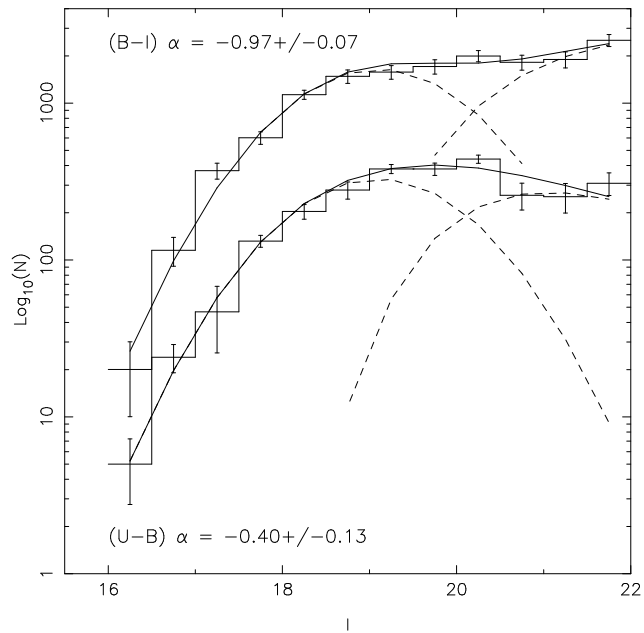}
\caption{The LF function for galaxies in the red elliptical sequence
in the c-m -- as identified independently from the \UB\ and \BI\  c-m
diagrams (Fig.~5).  Notice the relatively flat distribution 
at the faint end in the \UB\ sample, with an apparent
a dip in the counts fainter than $I\sim 20.5$ ($M_V=M^\ast + 2.5$).
The composite Gaussian+Schechter function fits to the distributions
are plotted as solid curves, with the individual components given
as dashed lines.  The two LF's have been offset vertically for clarity.}
\end{center}
\end{figure}

To search for the cause of the dip we select those galaxies lying along
the elliptical sequence in the \BI\--$I$ c-m plane and study the
distribution of their \UB\ colours.  We show in Fig.~10 the
distributions for two independent magnitude slices, $I=19.0$--20.5 and
$I=20.5$--22.0, which bracket the position of the dip.  The median
photometric errors for galaxies within these two samples are
$\delta_{(U-B)} = 0.08$ and $\delta_{(U-B)} = 0.19$.  We show the
observed distributions in \UB\ colours in Fig~10, the galaxies lying at
$(U-B)\le 0.1$ are those which correspond to the original definition of
``UV+''.  The two samples have intrinsic dispersions of $\delta_{(U-B)}
= 0.23$ and $\delta_{(U-B)} = 0.65$ after removing the contributions
from the photometric errors on the objects and the cluster-to-cluster
variations in mean colours.   Restricting the initial \BI\ selection to
cover a similar colour range for the samples brighter and fainter than
$I=20.5$ does not reduce the \UB\ colour range seen for the $I>20.5$
sample.  Further splitting the $I=20.5$--22.0 sample on the basis of
their \BI\ colours, we confirm that both halves show similar median
\UB\ colours as well as dispersions:  $<\! (U-B)\! >=0.31\pm 0.44$ and
$<\! (U-B)\! >=0.34\pm 0.47$ for the redder and bluer halves
respectively, indicating that there is no strong colour-colour
correlation within this sub-sample.  We therefore conclude that the red
galaxies fainter than $M_V \sim -17.7 + 5 \log h$ have a substantially
broader range of mid-UV colours than those galaxies brighter than this
limit.  

To identify morphologically  the faint, red cluster population we
turn to archival Hubble Space Telescope WFPC-2 imaging.
Unfortunately, such images are only available for a small region in one
of our clusters, A2390\footnote{Very shallow images ($\sim 0.3$~ks) of an area
of A1758 have been taken but these are useless for our purposes.}.  The A2390 data comprises 8.4~ks integration in F555W ($V$)
and 10.5~ks in F814W ($I$), they are discussed further in Kneib
et al.\ (1997).  Of the 78 objects with $20.5\le I\le 22.0$ which fall
on the \BI\ c-m sequence used for Fig.~10, only nine lie within the
WFPC-2 field, of these we would expect $1.6\pm 0.6$ to be field
contamination.  The morphologies of these galaxies were visually
determined using the scheme of Smail et al.\ (1997a).  The distribution
is 3 E's, 3 S0's, an S0/a, an Sc and a gravitationally lensed arc (the
giant arc in A2390).  The colours of the arc are at the blue end of the
distributions, especially in \UB, while the Sc galaxy is the reddest of
the nine galaxies in \BI, but the second bluest in \UB\ (after the
arc), indicating that it is likely to be background.  Ignoring these
field objects, the remaining seven galaxies split roughly equally into
S0 and E.  With only three galaxies in each class (we remove the S0/a),
our conclusions are limited, however, we note that the S0 galaxies tend
to be bluer in both \UB\ ($<\! (U-B)\!  >_{\rm S0}=0.58\pm 0.16$ versus
$<\! (U-B)\!  >_{\rm E}=0.77\pm 0.06$, where the errors on the means
are bootstrap estimates) and \BI\ ($<\! (B-I)\!  >_{\rm S0}=3.01\pm
0.08$ versus $<\!  (B-I)\! >_{\rm E}=3.18\pm 0.06$), as well as
spanning a wider range in \UB\ colours than the ellipticals,
$\sigma_{\rm S0} = 0.34$ as opposed to $\sigma_{\rm E} = 0.12$.  To
determine the significance of this difference in the dispersions, we
simulate a Gaussian distribution with a dispersion equal to that of the
combined E+S0 sample and ask how often we expect the ratio of the
relative dispersions of random subsets of three galaxies to exceed that
observed ($\sigma_{\rm S0}/\sigma_{\rm E} = 2.8$).  This occurs  in
only 2\% of cases, which would imply that there maybe a real 
difference between the \UB\ colour ranges of E and S0 galaxies.  
We conclude that the majority  of the cluster galaxies with $I\geq 20.5$
and red \BI\ colours are spheroidal systems (E or S0) and that the
bluer half of these contains 70\% S0 galaxies.

The relatively strong 4000\AA\ breaks in the UV+ galaxies means that in
the absence of dynamical effects, we would expect them to undergo only
modest fading in their restframe optical luminosities between $z\sim
0.24$ and today, not much above that expected for the elliptical
population, $\Delta M_V\sim -0.3$.  We therefore would predict that
this group would contribute $\sim 50$\% of the population with strong
4000\AA\ breaks at typical luminosities of around $M_V \sim -17.4 + 5
\log h$ in massive local clusters.  We compare the expected
characteristics of this population with those observed for different
local cluster galaxies in \S 4.

%
%
\begin{figure}
\begin{center}
\psfig{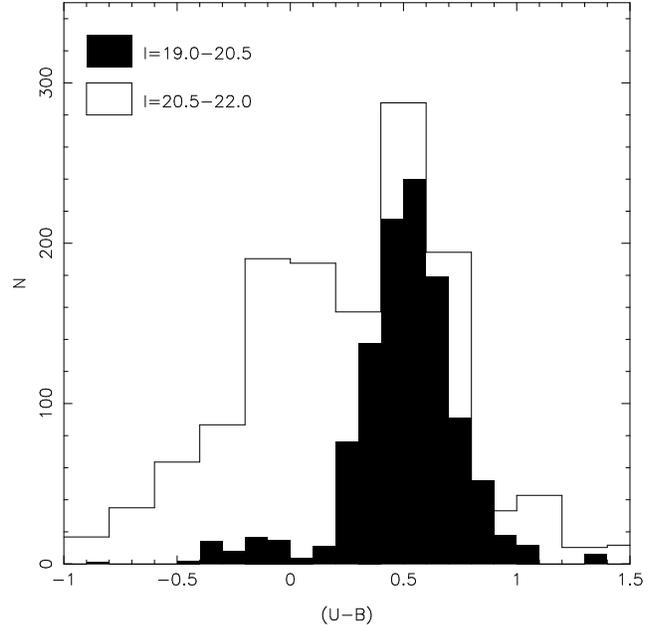}
\caption{The distribution in \UB\ colour for galaxies lying on the
elliptical sequence in the \BI\--$I$ c-m diagram.  This is shown for
two magnitude ranges, $I=19.0$--20.5 and $I=20.5$--22.0, which span the
position of the break in the \UB\ c-m sequence.  Note the blue wing to
colour distribution of the fainter sample, compared to the relatively
narrow distribution seen just brighter.  This wing contains $\sim 50$\%
of the population at fainter magnitudes. Both samples have been
selected using the same fixed-width envelope around the c-m sequence
and are corrected for the slope of the elliptical c-m sequence and
field contamination.  The brighter sample has been renormalised for the
purposes of comparison and all negative bins have been truncated.}
\end{center}
\end{figure}

\subsection{Blue Cluster Galaxies}

After the red spheroidal sequence, the next most prominent feature in
Figs.~3 and 5 is the population of faint, blue galaxies.  While some
proportion of these objects in Fig.~5 will be field galaxies, the
distributions shown in Fig.~3 have been corrected for this
contamination and hence the population of blue galaxies seen there is
associated with the cluster, although they do not show the strong
central concentration of the red population (Fig.~4).  

First we determine the blue fractions ($f_b$) in our clusters,
following the prescription of Butcher \& Oemler (1984, BO).  They
estimate the radius in the cluster which contains 30\% of the
population brighter than $M_V=-18.5 + 5 \log h$.  Using this radius
they then calculate the fraction of the $M_V\leq -18.5 + 5 \log h$
population within that aperture which have restframe $(B-V)$ colours
bluer than 0.2 mag below the elliptical sequence.  Defining the cluster
centre from our X-ray images, and correcting for the field counts, we
determine the values for $R_{30}$ and $N_{30}$, the number of cluster
galaxies within this radius, given in Table~5.  Table~5 also gives the
concentration index, $C=\log R_{60}/R_{20}$, as defined by BO.  The
only remaining step is to determine the equivalent colour boundary for
our passbands.  We have the colour of the elliptical sequence in each
cluster (\S 3.1), we need to transform an offset of $(B-V)=0.2$ in the
restframe into a difference in the observed \BI\ colour at $z\sim
0.24$.  For this purpose we simply fit the variation in restframe
$(B-V)$ colour with observed \BI\ colour for the non-evolved SEDs used
earlier.  ~From these fits we can derive the  \BI\ offset which
provides the equivalent colour difference in the restframe $(B-V)$.  A
colour of $(B-V)_{\rm E}-0.2$ is between an Sab and Sbc SED (0.67 Sab +
0.33 Sbc) and equivalent to an offset of $\Delta (B-I)= -0.58$ from the
observed elliptical c-m sequence.  Using this limit and the observed
elliptical sequences we determine the blue fractions given in Table~5.
We have one cluster in common with the original BO study: A1758.  For
this we derive $C=0.45$, $R_{30}=2.1$ arcmin and $f_b = 0.08\pm 0.06$,
in good agreement with the values published by BO:
$C=0.49$, $R_{30}=2.4$ arcmin and $f_b = 0.09\pm 0.04$.

%
%
\begin{figure}
\begin{center}
\psfig{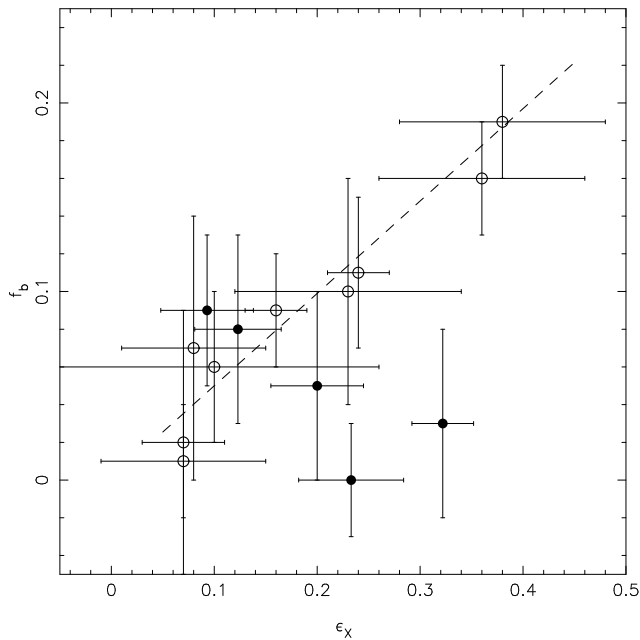}
\caption{A plot of the blue fraction of the cluster population, $f_b$,
versus the ellipticity of the cluster's X-ray emission measured within
a radius of 750 $h^{-1}$ kpc.  The points marked as open symbols come
from the analysis of Wang \& Ulmer (1997), the filled symbols are
from this work (we have used our measurements for A1758).  The dashed
line is the best fit relation to the data from Wang \& Ulmer (1997).
}
\end{center}
\end{figure}

The median $f_b$ for the concentrated clusters in our sample, those
with $C_{30}\geq 0.35$, is $<\! f_b\! > = 0.04\pm 0.02$.  This is
slightly lower than the  value of $<\! f_b\! > = 0.09\pm 0.04$ for a
similar sample of $z=0.2$--0.3 clusters in Butcher \& Oemler's original
work, perhaps indicating a tendency for these massive clusters
to contain a smaller proportion of active galaxies than typically seen
in a broad range of collapsed structures at their epochs.  It should be
a goal of any future studies to investigate such possibilities by contrasting
the properties of galaxy populations in massive and less massive
structures at a single epoch.  While the individual cluster cores
contain so few luminous, blue galaxies (typically 2--3 per cluster)
that the individual values of $f_b$ are not well determined, the
presence of such a small number of star-forming galaxies in the core
regions of these clusters does mean that they do not show any
Butcher-Oemler effect.

The scatter in our $C\geq 0.35$ sample is considerable, $\Delta f_b =
0.06$, and somewhat intriguing given that the clusters are all
concentrated and in addition span both a small range in mass and a
restricted redshift range.  Looking at the concentrated clusters with
the two highest $f_b$ values (we ignore the two $C\leq 0.35$ clusters)
we can see no obvious distinction between them and the remainder of the
sample.  Similarly focusing on the most bimodal cluster, A1758, we also see
no indication of atypical blue fractions in this system.  Evidence has
been presented for recent mergers in both A2219 (Smail et al.\ 1995b)
and A2390 (Pierre et al.\ 1996) from detailed lensing analysis,
although the blue populations in neither cluster shows any strong
evidence for a recent influx of star-forming galaxies.  We conclude
that the wide range in $f_b$ seen in our sample probably results from
small accretion events which, while adding a few star-forming galaxies
to the cluster, do not substantially alter its morphology, mass or
X-ray luminosity.  This is not to say that the accretion of a more
massive structure would not effect the cluster, just that such events
appear not to substantially alter the mix of galaxies in the cores of
massive clusters at $z\sim 0.2$--0.3.
 
In the light of this we address the recent claim by Wang \& Ulmer
(1997, WU) of a correlation between the blue fraction, $f_b$, and the
ellipticity of the X-ray emission on large scales in a sample of
clusters at $z\sim 0.15$--0.6.  We show the original data from Wang \&
Ulmer (1997) in Figure~11, where we also give similar observations of 5
clusters in our sample (those for which PSPC images are available).  We
have remeasured the ellipticities of the X-ray emission in all the
clusters, both those in WU and the new clusters added here, inside a
radius of  750 $h^{-1}$ kpc, to confirm that our new measurements are
on the same scale as the earlier values.  We find a mean offset between
the WU values and our measurements of:  $<\! \epsilon - \epsilon_{WU}\!
> = -0.004\pm 0.084$, with a typical deviation of $0.8\sigma_{WU}$
between the estimates.  We conclude that our ellipticity measurements
are in good agreement with those presented in WU and hence we can  add
our clusters to their sample.  The addition of our high luminosity
clusters to the already heterogenous sample from WU obviously does not
improve the correlation seen in their data (although neither does it
completely destroy it).  Moreover, we note that the presence of two
points at high--$\epsilon_X$ and high--$f_b$ appears to be responsible
for much of the significance of their original correlation.  One of
these is the cluster A2125, which has an X-ray luminosity of only a
tenth of the next lowest luminosity cluster in the sample, and as such
is clearly a very different object to the remaining clusters in the
figure.  Removing this single object to provide a more homogenous
sample also removes any statistically significant correlation between
$f_b$ and $\epsilon_X$.  We therefore caution that the
$f_b$--$\epsilon_X$ correlation reported by Wang \& Ulmer (1997) may be
simply the result of a small and diverse sample.  The lack of any
correlation between $f_b$ and $\epsilon_X$ for luminous X-ray clusters
at $z\ls 0.3$ should not be surprising given the generally low level of
Butcher-Oemler activity in these clusters and their wide range of
morphologies.

While our sample of clusters does lack a large population of luminous
blue galaxies, looking at the values in Table~2 we can see that they do
harbour a substantial population of star-forming galaxies, albeit at
considerably fainter magnitudes than those used to trace the
Butcher-Oemler effect.   As we show below the characteristic $V$-band
luminosity of the blue populations in these clusters is considerably
fainter than the cluster ellipticals, in contrast to the situation in
more distant clusters at $z\sim 0.5$ where the blue, spiral population
have a characteristic luminosity brighter than that of the ellipticals
(Smail et al.\ 1997a).  This may be hinting at a redshift--luminosity
relation for the active populations in the clusters, with the
characteristic luminosity of the star-forming class of galaxies
increasing with redshift.  A similar proposal has recently been made
for the evolution of the star-forming population in the distant field
(Lilly et al.\ 1995).    Lilly et al.\ (1995) claim that the
characteristic luminosity of the bluer field galaxies may brighten by
up to a magnitude out to $z\sim 0.5$.  Combining this evolution with
the increased accretion rate onto massive clusters at moderate
redshifts it may be possible to explain the redshift evolution of the
Butcher-Oemler effect using a simple infall model (Bower 1991).    

%
%
\begin{figure}
\begin{center}
\psfig{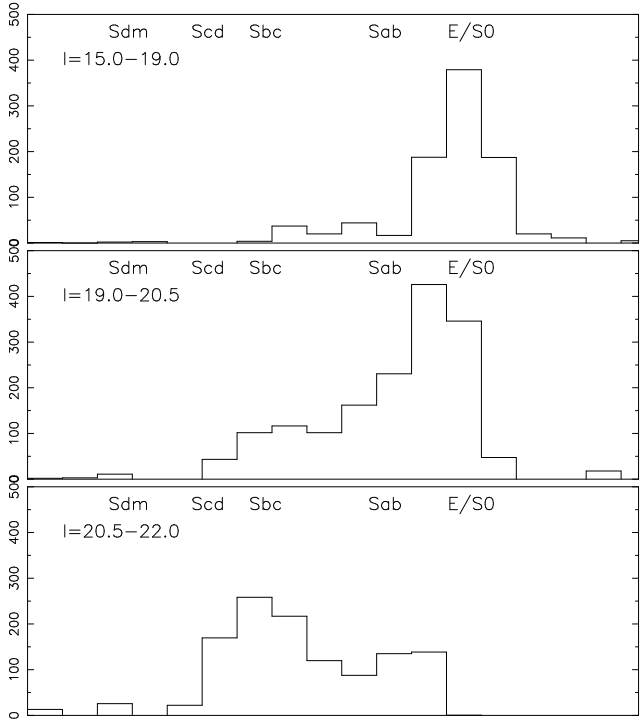}
\caption{The combined distribution of galaxies from the  10 clusters,
after field correction, projected along the axis defined by the colours
of the non-evolved SEDs shown in Fig.~3.  This is shown for each of the
three independent magnitude slices, $I=15.0$--$19.0$, $I=19.0$--$20.5$
and $I=20.5$--$22.0$ in Fig.~3.  We mark the equivalent spectral type
for $L^\ast$ galaxies across the tops of the figures.  No correction
has been made for the variation in elliptical colour with luminosity,
this results in a gradual shift of the elliptical sequence to the left
in the fainter samples.}
\end{center}
\end{figure}

We now look in more detail at the colour distribution within this faint
blue cluster population to better understand its subsequent evolution.
We address this using the two dimensional distributions from Fig.~3 and
project these onto the axis defined by the  non-evolved SEDs seen in
that figure.  This allows us to crudely classify the various galaxy
populations, in terms of their equivalent present-day spectral type.
We see a broad distribution of spectral classes, the frequency and
breadth of this distribution is at odds with that seen at similar
luminosities in comparably massive local clusters.   Fitting to the
luminosity distribution of the population with colours of an Sab or
bluer in \BI, we find $I^\ast = 20.5\pm 0.5$ and a steeply rising faint
end slope, $\alpha = -1.4\pm 0.3$.  Thus the characteristic luminosity
of this population is $M_V \sim -17.5 + 5 \log h$, about 2.5 mag
fainter than the elliptical population.  This is in contrast to more
distant clusters where the spiral population has a characteristic
luminosity similar to that of the cluster ellipticals (Smail et
al.\ 1997a).  Furthermore, the bulk of this population have colours
similar to present-day Sbc--Scd galaxies, with comparatively few as
blue as would be expected for vigorously star-forming Sdm galaxies.
Although it should also be kept in mind that the absence of objects as
blue as very late-type spirals from our clusters may be partly
explained by the limiting magnitude of our sample ($M_V \gs -16 + 5
\log h$).  With the advent of mid--UV surveys of local clusters (Brosch
et al.\ 1997), it will become possible to trace the evolution of the UV
luminosity function (and hence the mean star-formation rate) in cluster
environments from $z=0$ and compare this to the strong increase in star-formation seen in the distant field, amounting to a factor of $\sim 2
\times$ increase in the luminosity density  at 2800\AA\ out to $z\sim
0.2$ (Lilly et al.\ 1996).  Such a comparison would determine whether
the mean star-formation rates in clusters evolves in a similar manner
to that in the surrounding field, or whether the cluster environment
produces an additional long term decline in the star-formation of
cluster galaxies at recent epochs.

If the cluster environment is producing a widespread decline in the
star-formation of member galaxies, we can ask how the clusters we see
at $z\sim 0.24$ might appear today, in particular where the various
galaxy populations we have identified would appear in the luminosity
distribution of a local rich cluster.  To do this we need to model the
evolution in their star-formation rates with time. This can only be
done approximately, of course, given our lack of knowledge of their
detailed dynamical history and previous star-formation. Nevertheless,
at this stage it sufficies to determine what the simplest model for the
star-formation rate of the cluster population might predict at the
present day.   We start with the
expectation that the star-formation in all cluster galaxies will
decline towards the present, and hence for simplicity we have 
chosen to truncate the star-formation in all the cluster galaxies at
$z=0.24$ and evolve them forward to the present day, allowing the
galaxy's stellar populations to fade and  redden by the amount expected
from the models.  One attractive model for stopping star-formation in
cluster galaxies is to deplete their gas reservoirs by removing their
halos through interaction with the intracluster medium (Larson, Tinsley
\& Caldwell 1980).  While other more drastic alternatives are possible,
and may be necessary to form some of the more extreme spectral features
observed in some cluster galaxies (Couch et al.\ 1994), this basic
mechanism is sufficient to produce the drop in star-formation required
in our simple model.  We will next assume that a galaxy whose colours
are similar to a particular spectral type at $z=0.24$ would, if left
undisturbed, retain that spectral type until the present day.  This
allows us to use the evolutionary models of Fioc \& Rocca-Volmerange
(1997) which fit the local spectral types (E--Sdm) to describe the star-formation histories of the various galaxy populations in our clusters.
We can then take these models and truncate the star-formation at
$z=0.24$ before evolving them to $z=0$.  Using Fioc \&
Rocca-Volmerange's spectral synthesis code (PEGASE, Fioc \&
Rocca-Volmerange 1997) we can then determine the amount of fading and
reddening the various spectral types would undergo, in their restframe
$R$ and $(U-R)$ respectively.  The evolution of the stellar populations
after the star-formation (if any) is truncated produces different
degrees of subsequent reddening between $\Delta(U-R)=0.09$ for an
elliptical to $\Delta(U-R)=0.77$ for an Sdm, the equivalent fading in
the restframe $R$ is $\Delta R = 0.30$ for an elliptical and $\Delta R
= 0.94$ for an Sdm.\footnote{Here we assume a cosmology with $h=0.5$
and $\Omega=1$, giving a look-back time to $z=0.24$ of 4 Gyrs.}

The predicted fading and reddening from our model should be viewed as
the maximum allowed for a continously star-forming galaxy, both are
lower in the event that either  the star-formation was halted prior to
$z=0.24$ and we are observing the already fading remnant, or the
star-formation continues to lower redshift before being truncated.
However, the predicted fading and reddening can be exceeded if the star-formation is burst-like, as appears to be necessary for modelling some
of the spectral properties of distant cluster galaxies (e.g.\ Couch et
al.\ 1997).  More substantial fading of this population would also
result if dynamical processes (stripping, harassment, etc) removed
stellar material from the galaxies.  Moreover, we would expect
different fading rates across this population in those scenarios where
the efficiency of removing material is a function of bulge strength
(i.e.\ Hubble type), as appears to be the case for the harassment
mechanism (Moore et al.\ 1996, 1997).  Indeed, the harassment model
predicts substantial stripping from later-types (Sdm/Irr) and low
surface brightness galaxies, possibly amounting to $\sim 90--$95\% of
their stellar material, while leaving early-type Sab's relatively
untouched (apart from the truncation of the star-formation in their
disks).   Thus it is conceivable that the strong-bulge, early-type
spirals (Sab) would suffer only the fading expected from the stellar
population model, whereas the later-type spirals would  fade
considerably more than our predictions, possibly up to 2--4 mags (Moore
et al.\ 1997), due to the stripping of stars from these systems.
Better predictions from the various theoretical models are required
before we can make more detailed comparisons.  

To begin with we ignore possible dynamical effects, and simply apply
the evolution expected from our stellar models to the original
distribution of galaxies in the cluster, as a function of their
spectral type (Figure~12). We show the results of this in Figure~13,
where we give both the original distribution at $z=0.24$ and that which
would be observed for the same cluster at $z=0$.  As can be seen in
Figure~13, the truncation of all the star-formation in the clusters at
$z=0.24$ results in a large population of faint, mid-type spiral
galaxies moving towards the elliptical sequence.  Concentrating on
these fading spirals we would predict a characteristic magnitude of
$M_V\sim -17 + 5 \log h$ at the present day, and the population would
retain the steep faint end slope characteristic of the original blue
cluster population ($\alpha = -1.4\pm 0.3$).  Adding the
dynamically-driven, type-dependent fading discussed above would tend to
produce two families of remnents.  The stronger bulged, and typically
brighter, early-type spirals would end up around $M_V\sim -17 + 5 \log
h$, perhaps identifying them  as precursors to the UV+ group.  While
the bluer and later-type spirals would fade to $M_V\gs -16 + 5 \log h$,
providing a natural source of the large population of dwarf
elliptical (dE) galaxies seen at these magnitudes in local clusters
(e.g.\ BST; Gregory \& Thompson 1993).

For completeness we repeat the fit to the luminosity distribution of
galaxies lying along the red sequence predicted for the local clusters
(using our simple fading model).  We find that the evolution of the
blue population leads to a general steepening of the faint end of the
distribution.  Repeating the fit of a composite Gaussian+Schechter
function to the luminosity distribution along the red sequence, we find
that the faint end slope is expected to steepen by $\Delta \alpha =
-0.34$ from $\alpha \sim -1$ to closer to $\alpha \sim -1.4$ as a
result of the truncation of star-formation at $z=0.24$.  Here we have
evolved the parameters describing the brighter elliptical component to
reflect the expected evolution to $z=0$.  The resulting faint-end slope
lies close to that seen in the Coma cluster today, $\alpha = -1.4$
(Biviano et al.\ 1995).  While our simple model has provided a
possible outline of where the remnents of the
galaxy populations identified in the distant clusters might be found
today, we reiterate that much more detailed models, including information
on both the dynamics and the previous evolutionary history of the
galaxy populations, are needed to conclusively tackle this problem
(c.f.\ Baugh et al.\ 1996).

%
%
\begin{figure}
\begin{center}
\psfig{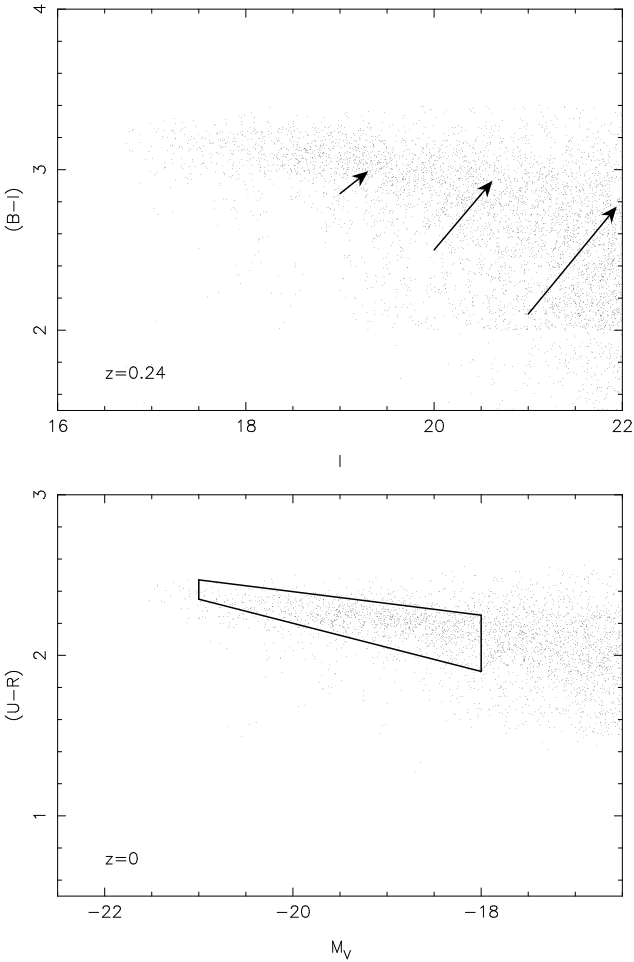}
\caption{The effect on the observed c-m diagram of terminating SF in
all the cluster galaxies at $z=0.24$.  The upper panel shows the
observed c-m distribution in \BI\ at $z\sim 0.24$ (roughly restframe
$(U-R)$), this has been statistically corrected for field contamination
using nearest-neighbour removal. The vectors indicate the extent of
reddening and fading of the galaxies in our model 
by the present day if their star-formation was truncated at $z=0.24$ (for an Sab, Scd and an Sdm).  The
lower panel shows the distribution in $(U-R)$ colour for the cluster
population as it would be seen at $z=0$, the effects of the fading and
reddening of the stellar populations in the galaxies is clearly
visible.  We also show on the lower panel the region of the 
c-m plane populated by the E and S0 galaxies in Coma studied in BLE.}
\end{center}
\end{figure}

\section{Discussion and Conclusions}

We have analysed the galaxy populations in the 10 luminous X-ray clusters in
our sample  on an individual cluster-by-cluster basis.  The interest
of this sample comes from its selection from the most massive clusters
at $z\sim 0.2$, allowing a simple comparison to the properties of the
richest local clusters.  Furthermore, using the uniformity of our
sample and dataset we have been able to combine the galaxy populations
across all the clusters to provide a more detailed view of their
typical populations.  The galaxy mix in our clusters at bright
magnitudes is characterised by a population of luminous red galaxies,
with only a small fraction of blue, star-forming galaxies ($\sim 5$\%)
and an even smaller fraction of UV+ objects ($\sim 0.5$\%). We find
that the typical colours of the luminous ($M_V \leq -18.5 + 5 \log h$)
red galaxies are highly homogenous across all the clusters in our
sample.  The intrinsic scatter in the typical colours of this
population between clusters is $\ls 2$\% in restframe $(U-R)$.  In our
bluer colour, restframe $(2900$\AA$-U)$, the scatter may be larger
although our photometric precision is not as high, leading to a limit on
the cluster-to-cluster variation of $\ls 5$\% in their mean colours.
These observations thus extend the studies of BLE and Ellis et
al.\ (1997) across a larger sample of clusters and into the mid--UV.
We conclude that the brightest cluster galaxies comprise a homogeneous
population longward of 2900\AA\ in their restframe.  We illustrate the
type of constraints on cluster formation which this observation can
provide using a very simple model.  The model allows us to convert the
observed cluster-to-cluster scatter in the mean colours of cluster
ellipticals into a joint constraint on the lower limit of the formation
epoch for this population and the maximum size of the structures in
which they can coherently form.  While the model we use is not
physically well motivated, it does provide some insight into the
information on cluster formation which these observations can provide.
A rigorous theoretical analysis of the scatter expected in the
colours of luminous ellipticals both within clusters (e.g.\ Ellis et
al.\ 1997) and between clusters is urgently required.

A comparison between the integrated luminosity of the red cluster
galaxies with the X-ray temperatures of 7 of the clusters in
our sample shows a good correlation between in the two observables
with a scatter of only $\sim 17$\%.  We conclude
that the total luminosity of the elliptical population in the clusters is
a reasonable tracer of the cluster mass.  Again this is an interesting
observation to tackle with theoretical models of cluster formation
and growth, which can be robustly applied to our sample due to
the well-defined cluster selection originally used.

The picture of wide spread uniformity abruptly changes as we probe
fainter in the clusters.  Beyond $M_V \geq -17.5 + 5 \log h$ we start
to detect an increasing population of blue cluster galaxies.
These objects fall into two groups, galaxies blue in both \UB\ and
\BI\ (about $\sim 30$\% of the population) and those which while having
blue \UB\ colours are red in \BI: the UV+ galaxies, these are $\sim 5$\%
of the total cluster population and comprise $\sim 50$\% of the faint cluster
population with red \BI\ colours.  The latter class of galaxies defines
an extension of the narrow red sequence of the luminous ellipticals in
the \BI\ c-m plane.  The same is not true in \UB\ where they exhibit a
substantial spread in colours, in contrast to the brighter
ellipticals.  A galaxy's flux around 2900\AA, as probed by our
\UB\ colours, traces residual star-formation (Dorman et al.\ 1995),
being $\sim 50$\% more sensitive to low levels of star-formation than
our \BI\ colour for standard IMFs.  Thus the wide range in \UB\ colours
for this population indicates a spread in current star-formation,
albeit at a low enough level that it doesn't substantially perturb
their red \BI\ colours.  The lack of a strong signature in the
4000\AA\ colours of these galaxies means they are not included in the
standard Butcher-Oemler definition for the blue fraction of the cluster
population, although neither are they completely passive.

The fates of these two blue populations in present-day clusters are of
considerable interest.  We propose that, in the absence of dynamical
processes or a reactivation of their star-formation, the UV+ population
will undergo only modest fading in their optical luminosities from that
observed at $z\sim 0.24$.   As the clusters we study are amongst the
most massive formed at their epoch, we assume that their galaxy
populations should evolve to be similar to that seen in the richest
local clusters.  Therefore,  these galaxies should contribute a large
proportion of the quiescent, red $M_V \sim -17.4 + 5 \log h$ population
in the most massive local clusters ($M\geq M_{\rm Coma}$).  Of the
$I\geq 20.5$ galaxies in the distant clusters with red \BI\ colours,
roughly 50\% lie in a blue tail in \UB.  The high proportion of such
objects indicates that a large fraction of the quiescent $M_V \sim
-17.4 + 5 \log h$ population in local clusters must have passed through
this stage.   The morphological mix in the Coma cluster at these
magnitudes is dominated by S0 galaxies, 65\% of the population, with
ellipticals making up 30\% and Sa's most of the remaining 5\% (Dressler
1980).  Thus while some of the UV+ galaxies may continue to form stars
at a low level and hence appear as Sa galaxies in the local clusters,
the majority must be precursors to the S0 or E populations.  Moreover,
the typical magnitude of these galaxies corresponds to a luminosity
close to the characteristic value for the S0 populations in local
clusters.  BST claim a mean luminosity of $<\! M_V\! >  \sim -17.3 + 5
\log h$ for the S0 galaxies in Virgo (assuming $(B-V)\sim 0.6$), while
Thompson \& Gregory (1993) find $<\!  M_V\! >  \sim -17.8 + 5 \log h$
for the S0 population of Coma.  In the light of this and the
observations of a rapid decline in the numbers of S0 galaxies in $z\gs
0.4$ clusters (Dressler et al.\ 1997), we suggest that the faint
galaxies observed in our $z\sim 0.2$--0.3 clusters with red
\BI\ colours, but blue \UB\, may be the precursors of today's S0
cluster population, whose characteristics they would closely match. 

By morphologically classifying a small number of the $I\geq 20.5$ Red
and UV+ population using WFPC-2 imaging of A2390 we demonstrate that
the faint, red cluster population does consist of spheroidal galaxies
and we also show marginal evidence that the S0 component of this
population may exhibit a wider range, and typically bluer, mid--UV
colours than the faint ellipticals.  The substantial deficit of S0s in
clusters at $z\sim 0.5$ (Dressler et al.\ 1997), along with our
suggested identification of the UV+ population with the precursor's of
the S0 population of local rich clusters, would indicate that $z\sim 0$
S0s should show some signs of their recent transformation.   The
spectral signatures of recent star-formation have been seen in samples
of spheroidal galaxies from the Coma cluster (Caldwell et al.\ 1993),
although these have not been linked specifically to the S0 population.
Moreover, in a comparison of the dispersions in the $(U-V)$ colours of
luminous ellipticals and S0s in Coma, BLE concluded that both
populations  showed essential no intrinsic scatter.  Clearly issues of
luminosity and morphology must be addressed before we can fully
understand the evolution of the galaxy populations of clusters.
Central to this understanding will be further work on classifying
samples of faint galaxies in $z\sim 0.1$--0.3 clusters, these will
enable us to robustly track the morphological evolution of this
population (Couch et al.\ 1997a). 

The fate of the blue, star-forming galaxies in these distant clusters
is more speculative.  These are typically low luminosity systems ($L_V
\sim 0.001 L^\ast$), exhibiting relatively modest star-formation.  We
illustrate one possible evolutionary pathway, which the population
would follow if their star-formation was terminated at $z=0.24$, this
produces a population of faint, red galaxies in local clusters with a
steep luminosity distribution.  These characteristics are similar to
those of the dwarf elliptical population which dominates local clusters
at faint magnitudes (BST).   Recent spectroscopic observations of
similar objects in a cluster at $z=0.4$ has shown them to
have very low masses (Koo et al.\ 1997) and so we caution that these
low luminosity galaxies may be be quite fragile and hence the cluster
environment could have a significant impact on their luminosity and
morphological evolution through dynamical processes (Moore et
al.\ 1996).  If substantial amounts of baryonic material are removed
from late-type galaxies during their accretion onto the clusters we
would expect this material to be deposited into the cluster potential
as either intracluster gas or intracluster light.  Tracing the
evolution of these components within rich clusters at $z\sim 0$--0.5
may provide the most direct test of the stripping and harassment
mechanisms (Moore et al.\ 1997).

In conclusion, there is a growing body of evidence which indicates that
the passive red galaxy populations in local clusters are produce by a
diverse range of processes.  The most luminous elliptical galaxies
appear to have formed at very early epochs ($z\gs 3$, Ellis et
al.\ 1997), while the bulk of the S0 population have come into being
much more recently ($z\ls 0.5$, Dressler et al.\ 1997) as may the lower
luminosity dwarf ellipticals (Koo et al.\ 1997).   By combining the
galaxy samples from our uniform survey of clusters we have identified a
class of galaxies in our distant clusters which would have
characteristics similar to those of the S0 population at low redshift,
but which still  show residual traces of star-formation, consistent with
their being previously more active.  Determining the morphologies of a
large sample of these objects in intermediate redshift clusters using
HST will provide a strong test of their relationship to both the
Butcher-Oemler populations of distant clusters and the local S0s
(c.f.\ Larson, Tinsley \& Caldwell 1980).  

\section*{Acknowledgements}

We would like to thank Steve Allen, Harald Ebeling, Andy Fabian and
Hans B\"ohringer for their work in defining the X-ray sample on which
our study is based.  We also thank Jim McCarthy and Jim Westphal for
their work in providing an efficient UV imaging capability on the 5-m,
essential for our observations, as well as Jim Schombert for kindly
loaning us his $U$ filter, David Hogg for providing his deep $U$ field
exposure, Alan Dressler for donating 5-m time to undertake the $BI$
imaging of the same field and David Buote for the analysis of the
shapes of the cluster X-ray emission used in Fig.~11.  We thank the
referee, Phil James, for his thorough reading of the paper and
constructive comments which improved the text, particularly in regard
to the discussion of the models.  Finally, we thank Rebecca Bernstein,
Richard Bower, Julianne Dalcanton and Roger Davies for extensive
discussions on the nature and evolution of cluster galaxies.  IRS
acknowledges support from a PPARC Advanced Fellowship and ACE from a
Royal Society Fellowship.  RDB acknowledges NSF grants AST 92-23370 and
AST 95-29170.  The research published here used the STARLINK computer
resources at the University of Durham.


\newpage

%
%

\begin{table*}
\caption{The log of observations and our cluster sample. We list the
field identification, position and cluster redshift, as well as the
total integrations times and seeing of the $UBI$ images of the
clusters.  We then give the 80\% completeness of the $I$ band
catalogues from comparison with deeper field counts and the estimated
reddening in these directions.  The final three columns give the
angular scale at the cluster redshift for our adopted cosmology (in
units of $h^{-1}$~kpc/$''$), the total area covered by our images in
each cluster in sq.\ arcmin,  the X-ray luminosity of the clusters
in the 0.2--2.4 keV band in units of $h^{-2}$ 10$^{44}$ ergs s$^{-1}$
and the X-ray temperatures in keV from Mushotzky \& Scharf (1997) where
available.}
\vspace{0.5cm}
\hspace{-1.3truecm}\hbox{\begin{tabular}{lccccccccccccccr}
\noalign{\medskip}
\hline\hline
\noalign{\smallskip}
{ID} &  {RA} &   {Dec}   & $z$ &  \multispan3{\hfil ~T$_{\rm exp}$~(ks)\hfil } & \multispan3{\hfil FWHM~($''$)\hfil } & $I$    & 
$E(B-V)$ & Scale & Area & $L_X$ & $T_X$ \cr
     & (J2000) & (J2000) &    &  $U$ & $B$ & $I$                & $U$ & $B$ & $I$  & (80\%) & &  &  &  \cr
\hline
\noalign{\smallskip}
A1682  & 13~06~52.4 & $+$46~33~05 & 0.226 & 3.0 & 0.6 & 0.5 & 1.19 & 1.47 & 0.87 & 22.5 & 0.03 & 2.30 & 92.55 & 2.79 & ---~\cr
A1704  & 13~14~24.5 & $+$64~34~30 & 0.220 & 3.6 & 0.5 & 0.5 & 1.37 & 1.63 & 1.31 & 22.5 & 0.04 & 2.26 & 89.30 & 1.69 & ~4.5\cr
A1758  & 13~32~44.0 & $+$50~32~33 & 0.280 & 3.0 & 0.5 & 0.5 & 1.17 & 1.27 & 1.05 & 22.5 & 0.02 & 2.64 & 86.46 & 2.81 & 10.2\cr
A1763  & 13~35~18.5 & $+$40~59~46 & 0.228 & 3.0 & 0.5 & 0.5 & 1.22 & 1.41 & 1.07 & 22.5 & 0.02 & 2.31 & 90.96 & 3.62 & ~9.0\cr
A1835  & 14~01~02.2 & $+$02~52~43 & 0.253 & 3.0 & 0.5 & 1.0 & 1.43 & 1.72 & 1.14 & 22.5 & 0.05 & 2.47 & 82.74 & 9.59 & ~8.2\cr
Zw7160 & 14~57~14.9 & $+$22~20~35 & 0.256 & 3.0 & 0.6 & 1.0 & 1.23 & 1.67 & 1.10 & 23.0 & 0.07 & 2.49 & 84.84 & 3.37 & ~5.5\cr
A2146  & 15~56~11.5 & $+$66~21~30 & 0.234 & 3.0 & 0.5 & 0.5 & 1.31 & 1.24 & 1.15 & 23.0 & 0.06 & 2.35 & 92.55 & 2.15 & ---~\cr
A2219  & 16~40~20.5 & $+$46~42~29 & 0.228 & 3.7 & 1.0 & 0.5 & 1.54 & 1.25 & 1.10 & 22.5 & 0.04 & 2.31 & 92.60 & 4.95 & 11.8\cr
A2261  & 17~22~26.8 & $+$32~07~59 & 0.225 & 3.0 & 0.5 & 0.5 & 1.35 & 1.29 & 1.04 & 22.5 & 0.07 & 2.29 & 93.01 & 4.51 & ---\cr
A2390  & 21~53~36.8 & $+$17~41~46 & 0.233 & 3.0 & 0.5 & 0.9 & 1.43 & 1.37 & 1.13 & 22.5 & 0.14 & 2.34 & 92.69 & 5.31 & ~8.9\cr
Field  & 00~53~23.2 & $+$12~33~58 &  --- & 28.0 & 1.0 & 1.0 & 1.10 & 1.20 & 1.10 & 23.0 & 0.00 & ---  & 81.00 & ---  & ---~\cr
\noalign{\smallskip}
\noalign{\hrule}
\noalign{\smallskip}
\end{tabular}}
\end{table*}
\bigskip

%
%
\begin{table*}
\caption{The relative proportions of the
various cluster populations we identify, for two absolute magnitude limits.
``Red'' galaxies have \UB,\BI\ colours of [0.1:0.9,2.9:3.6],
``Blue'' have [$-$1.1:0.9,1.9:2.9] and ``UV+''
[$-$1.1:0.1,2.9:3.6].  These values are for a fiducial cluster at
$z=0.24$ and refer to the whole area imaged in each cluster. 
The relative proportions have been corrected for field contamination.}
\vspace{0.5cm}
\begin{tabular}{lrrrrr}
\noalign{\medskip}
\hline\hline
\multispan6{\hfil $M_V\leq -18.5 + 5 \log h$\hfil }\cr
\noalign{\smallskip}
{ID} &  $N_{\rm red}$ &  $N_{\rm blue}$ & $N_{\rm UV+}$ & $f_{\rm blue}$ & $f_{\rm UV+}$ \cr
 \hline
\noalign{\smallskip}
A1682  &  133.0 &    2.0 &    1.7 &    0.01 &    0.01 \cr
A1704  &   69.6 & $-$4.4 &    1.8 & $-$0.07 &    0.03 \cr
A1758  &  160.1 &   20.1 &    1.9 &    0.11 &    0.01 \cr
A1763  &  143.3 & $-$3.7 & $-$1.2 & $-$0.03 & $-$0.01 \cr
A1835  &  148.7 &   11.7 &    5.0 &    0.07 &    0.03 \cr
Zw7160 &   69.3 &   12.3 &    0.9 &    0.15 &    0.01 \cr
A2146  &   66.0 &    0.0 & $-$1.3 &    0.00 & $-$0.02 \cr
A2219  &  180.0 &    8.0 & $-$0.3 &    0.04 & $-$0.00 \cr
A2261  &  131.9 &   12.9 & $-$1.3 &    0.09 & $-$0.01 \cr
A2390  &  105.0 & $-$5.0 & $-$1.3 & $-$0.05 & $-$0.01 \cr
\noalign{\smallskip}
All    & 1206.9 &   53.9 &    5.9 &    0.04 &    0.01 \cr
\noalign{\medskip}
\multispan6{\hfil $M_V\leq -17.0 + 5 \log h$\hfil}\cr
\noalign{\smallskip}
\hline
\noalign{\smallskip}
A1682  &  258.9 & 135.7 &  47.3 & 0.31 & 0.11 \cr
A1704  &  131.0 &  45.6 &   7.8 & 0.25 & 0.04 \cr
A1758  &  225.0 & 127.1 &   8.2 & 0.35 & 0.02 \cr
A1763  &  247.4 &  59.1 &  14.5 & 0.18 & 0.05 \cr
A1835  &  224.4 & 128.5 &  45.7 & 0.32 & 0.11 \cr
Zw7160 &  108.6 & 104.6 &   5.4 & 0.48 & 0.02 \cr
A2146  &   94.9 &  35.7 &   1.3 & 0.27 & 0.01 \cr
A2219  &  307.8 &  83.7 &  13.3 & 0.21 & 0.03 \cr
A2261  &  184.7 &  79.3 &   0.2 & 0.30 & 0.00 \cr
A2390  &  118.8 &  27.6 &   0.3 & 0.19 & 0.00 \cr
\noalign{\smallskip}
All    & 1901.5 & 826.9 & 144.0 & 0.29 & 0.05 \cr
\noalign{\smallskip}
\noalign{\hrule}
\noalign{\smallskip}
\end{tabular}
\end{table*}
\bigskip

%
%
\begin{table*}
\caption{The coefficients of linear fits to the \UB--$I$ and \BI--$I$ c-m for
red cluster galaxies.  The intercept values are evaluated at a magnitude 
equivalent to $I=18$ in a $z=0.24$ cluster.}
\vspace{0.5cm}
\begin{tabular}{lcccccc}
\noalign{\medskip}
\hline\hline
\noalign{\smallskip}
{ID} &  $z$ &   $A_{(U-B)}$ & $\delta A_{(U-B)}$ & $B_{(U-B)}$ & $\delta B_{(U-B)}$ & $N$ \cr
\hline
\noalign{\smallskip}
A1682  & 0.226 & 0.473 & 0.017 & $-$0.021 & 0.016 & 177 \cr
A1704  & 0.220 & 0.399 & 0.020 & $-$0.026 & 0.018 & 107 \cr
A1758  & 0.280 & 0.588 & 0.017 & $-$0.070 & 0.017 & 206 \cr
A1763  & 0.228 & 0.500 & 0.015 & $-$0.055 & 0.014 & 209 \cr
A1835  & 0.253 & 0.444 & 0.018 & $-$0.032 & 0.018 & 200 \cr
Zw7160 & 0.256 & 0.370 & 0.024 & $-$0.031 & 0.026 & 126 \cr
A2146  & 0.234 & 0.493 & 0.024 & $-$0.044 & 0.025 & 100 \cr
A2219  & 0.228 & 0.501 & 0.013 & $-$0.018 & 0.012 & 230 \cr
A2261  & 0.225 & 0.456 & 0.016 & $-$0.044 & 0.015 & 197 \cr
A2390  & 0.233 & 0.638 & 0.019 & $-$0.075 & 0.018 & 198 \cr
\noalign{\smallskip}
\noalign{\hrule}
\noalign{\smallskip}
\end{tabular}
\vspace{0.5cm}
\begin{tabular}{lcccccr}
\noalign{\medskip}
\hline\hline
\noalign{\smallskip}
{ID} &  $z$ &   $A_{(B-I)}$ & $\delta A_{(B-I)}$ & $B_{(B-I)}$ & $\delta B_{(B-I)}$ & $N$ \cr
 \hline
\noalign{\smallskip}
A1682  & 0.226 & 3.225 & 0.021 & $-$0.062 & 0.019 & 173 \cr
A1704  & 0.220 & 3.048 & 0.020 & $-$0.073 & 0.018 & 100 \cr
A1758  & 0.280 & 3.088 & 0.019 & $-$0.077 & 0.021 & 216 \cr
A1763  & 0.228 & 3.085 & 0.018 & $-$0.100 & 0.018 & 202 \cr
A1835  & 0.253 & 3.098 & 0.016 & $-$0.046 & 0.016 & 195 \cr
Zw7160 & 0.256 & 3.073 & 0.031 & $-$0.044 & 0.031 & 109 \cr
A2146  & 0.234 & 3.105 & 0.020 & $-$0.092 & 0.020 &  89 \cr
A2219  & 0.228 & 3.138 & 0.015 & $-$0.049 & 0.014 & 221 \cr
A2261  & 0.225 & 3.034 & 0.013 & $-$0.067 & 0.013 & 163 \cr
A2390  & 0.233 & 3.121 & 0.020 & $-$0.066 & 0.019 & 194 \cr
\noalign{\smallskip}
\noalign{\hrule}
\noalign{\smallskip}
\end{tabular}
\end{table*}
\smallskip

%
%
\begin{table*}
\caption{The intercepts of fits to the c-m relations in the clusters
adopting a fixed slope in each passband ($B_{(U-B)}=-0.041$ and $B_{(B-I)}=-0.067$). 
These are shown for the two magnitude-limited samples.
The intercept values are evaluated at a magnitude 
equivalent to $I=18$ in a $z=0.24$ cluster.}
\vspace{0.5cm}
\begin{tabular}{lcccccc}
\noalign{\medskip}
\hline\hline
\multispan7{\hfil $M_V\leq -20.0 + 5 \log h$\hfil }\cr
\noalign{\smallskip}
{ID} &  $A_{(U-B)}$ & $\delta A_{(U-B)}$  & $N_{(U-B)}$ & $A_{(B-I)}$ & $\delta A_{(B-I)}$ & $N_{(B-I)}$ \cr
 \hline
\noalign{\smallskip}
A1682  & 0.472 & 0.022 & 37 & 3.219 & 0.019 & 31 \cr
A1704  & 0.389 & 0.028 & 27 & 3.085 & 0.022 & 21 \cr
A1758  & 0.606 & 0.016 & 50 & 3.133 & 0.014 & 43 \cr
A1763  & 0.490 & 0.022 & 42 & 3.135 & 0.019 & 36 \cr
A1835  & 0.440 & 0.018 & 44 & 3.114 & 0.014 & 36 \cr
Zw7160 & 0.379 & 0.032 & 34 & 3.140 & 0.032 & 17 \cr
A2146  & 0.485 & 0.023 & 27 & 3.098 & 0.017 & 26 \cr
A2219  & 0.486 & 0.015 & 51 & 3.170 & 0.017 & 42 \cr
A2261  & 0.458 & 0.017 & 52 & 3.057 & 0.014 & 45 \cr
A2390  & 0.660 & 0.021 & 47 & 3.150 & 0.017 & 38 \cr
\noalign{\medskip}
\multispan7{\hfil $M_V\leq -18.5 + 5 \log h$\hfil}\cr
\noalign{\smallskip}
\hline
\noalign{\smallskip}
A1682  & 0.487 & 0.013 & 177 & 3.210 & 0.009 & 173 \cr
A1704  & 0.423 & 0.016 & 107 & 3.086 & 0.009 & 100 \cr
A1758  & 0.557 & 0.013 & 206 & 3.125 & 0.009 & 216 \cr
A1763  & 0.481 & 0.010 & 209 & 3.115 & 0.007 & 202 \cr
A1835  & 0.444 & 0.013 & 200 & 3.133 & 0.007 & 195 \cr
Zw7160 & 0.373 & 0.016 & 126 & 3.125 & 0.012 & 109 \cr
A2146  & 0.475 & 0.016 & 100 & 3.112 & 0.012 & ~89 \cr
A2219  & 0.506 & 0.011 & 230 & 3.171 & 0.006 & 221 \cr
A2261  & 0.446 & 0.014 & 197 & 3.068 & 0.008 & 163 \cr
A2390  & 0.614 & 0.013 & 198 & 3.149 & 0.010 & 194 \cr
\hline
\noalign{\smallskip}
\noalign{\smallskip}
\noalign{\hrule}
\noalign{\smallskip}
\end{tabular}
\end{table*}
\bigskip

%
%

\begin{table*}
\caption{The structural parameters of the clusters determined from
our galaxy catalogues.}
\vspace{0.5cm}
\begin{tabular}{lccccr}
\noalign{\medskip}
\hline\hline
\noalign{\smallskip}
{ID} &  C & $R_{30}$ & $N_{30}$ & $L_V$ & ~~$f_b$\hfil~ \cr
   &   & arcmin & & $h^{-2} 10^{10} L_\odot$ \cr
 \hline
\noalign{\smallskip}
A1682      & 0.32 & 1.75 & 45 & $368\pm 30$ & $-0.01\pm 0.05$ \cr
A1704      & 0.63 & 1.18 & 21 & $243\pm 24$ &  $0.10\pm 0.10$ \cr
A1758$^1$  & 0.45 & 2.10 & 65 & $536\pm 54$ &  $0.08\pm 0.06$ \cr
A1763      & 0.44 & 1.75 & 53 & $432\pm 32$ &  $0.03\pm 0.05$ \cr
A1835      & 0.42 & 1.37 & 52 & $487\pm 40$ &  $0.00\pm 0.03$ \cr
Zw7160     & 0.37 & 2.22 & 36 & $204\pm 15$ &  $0.08\pm 0.10$ \cr
A2146$^2$  & 0.45 & 0.80 & 14 & $246\pm 21$ & $-0.07\pm 0.02$ \cr
A2219      & 0.40 & 1.56 & 60 & $470\pm 23$ &  $0.08\pm 0.05$ \cr
A2261      & 0.32 & 2.13 & 66 & $424\pm 30$ &  $0.20\pm 0.08$ \cr
A2390      & 0.38 & 1.37 & 49 & $441\pm 29$ &  $0.05\pm 0.05$ \cr
\noalign{\smallskip}
\noalign{\hrule}
\noalign{\smallskip}
\end{tabular}
\bigskip
\begin{tabular}{cl}
1) & C, $R_{30}$, $N_{30}$ and $f_b$ calculated using S.E.\ component as centre.  \cr
2) & C, $R_{30}$, $N_{30}$ and $f_b$ calculated using the mean of the position of the two bright
galaxies as centre. \cr
\end{tabular}
\end{table*}
\bigskip


\begin{thebibliography}{}

\bibitem[Allen 1995]{swa95}
Allen, S.W., 1995, MNRAS, 276, 947.

\bibitem[Allen et al.\ 1997]{swa97}
Allen, S.W., et al., 1997, MNRAS, in press.

\bibitem[Arag\'on-Salamanca et al.\ 1993]{aetal93}
Arag\'on-Salamanca A., Ellis R.S., Couch W.J., Carter D., 1993, MNRAS,
262, 764.

\bibitem[Barger et al.\ 1996]{b96}
Barger, A.J., Arag\`on-Salamanca, A., Ellis, R.S., Couch, W.J., Smail, I.\
\& Sharples, R.M., 1996,  MNRAS, 279, 1.

\bibitem[Barger et al.\ 1997]{b97}
Barger, A., Arag\`on-Salamanca, A., Smail, I.,
Ellis, R.S., Couch, W.J., Dressler, A., Oemler, A., Butcher, H.\ \&
Sharples, R.M.,  1997, in prep.

\bibitem[Baugh et al.\ 1996]{b96}
Baugh, C.M., Cole, S.\ \& Frenk, C.S., 1996, MNRAS, 283, 1361.

\bibitem[Baugh et al.\ 1997]{b97}
Baugh, C.M., Cole, S.\ \& Frenk, C.S., 1997, MNRAS, in press.

\bibitem[Belloni et al.\ 1995]{bel95}
Belloni, P., Bruzual, A.G., Thimm, G.J.\ \& Roser, H.-J., 1995, A\&A, 297, 61.

\bibitem[Bertin \& Arnouts 1996]{ba96}
Bertin E.\ \& Arnouts, S., 1996, A\&A Supp., 117, 393.

\bibitem[Bessell 1990]{b90}
Bessell, M., 1990, PASP, 102, 1181.

\bibitem[Binggeli, Sandage \& Tammann 1985]{bst85}
Binggeli, B., Sandage, A.\ \& Tammann, G.A., 1985, AJ, 90, 1681. (BST)

\bibitem[Biviano et al.\ 1995]{bv95}
Biviano, A., Durret, F., Gerbal, D., Fevre, O.L., Lobo, C.,
Mazure, A.\ \& Sleazak, E., 1995, A\&A, 297, 610.

\bibitem[Bower 1991]{b92}
Bower, R.G., 1991, MNRAS, 248, 332.

\bibitem[Bower, Lucey \& Ellis 1992]{ble}
Bower, R.G., Lucey, J.R.\ \& Ellis, R.S., 1992, MNRAS, 254, 601. (BLE)

\bibitem[Broadhurst et al.\ 1997]{tgb97}
Broadhurst, T., Villumsen, J.V., Smail, I.\ \& Charlot, S.,
1997, ApJ, submitted.

\bibitem[Brosch et al.\ 1997]{b97}
Brosch, N., Formiggini, L., Almoznino, E., Sasseen, T.,
Lampton, M.\ \& Bowyer, S., 1997, ApJ, in press. 

\bibitem[Butcher \& Oemler 1978]{bo78}
Butcher, H.\ \& Oemler, A., 1978, ApJ, 226, 559.

\bibitem[Butcher \& Oemler 1984]{bo84}
Butcher, H.\ \& Oemler, A., 1984, ApJ, 285, 426.

\bibitem[Byrd \& Valtonen 1990]{bv90}
Byrd, G.\ \& Valtonen, G., 1990, ApJ, 350, 89.

\bibitem[Caldwell et al.\ 1993]{c93}
Caldwell, N., Rose, J.A., Sharples, R.M., Ellis, R.S.\ \& Bower, R.G.,
1993, AJ, 106, 473.

\bibitem[Colless 1989]{mcc89}
Colless, M.M., 1989, MNRAS, 237, 799.

\bibitem[Couch \& Sharples 1987]{cs87}
Couch, W.J.\ \& Sharples, R.M., 1987, MNRAS, 229, 423.

\bibitem[Couch et al.\ 1994]{cess}
Couch, W.J., Ellis, R.S., Sharples, R.M.\ Smail, I., 1994, ApJ, 430, 121.

\bibitem[Couch et al.\ 1997a]{c97a}
Couch, W.J., Barger, A.J., Smail, I., Ellis, R.S.\ \&
Sharples, R.M., 1997a, ApJ, submitted.

\bibitem[Couch et al.\ 1997b]{c97b}
Couch, W.J., et al., 1997b, in prep.

\bibitem[Dorman et al.\ 1995]{d95}
Dorman, B., O'Connell, R.W.\ \& Rood, R.T., 1995, ApJ, 442, 105.

\bibitem[Dressler 1980]{d80}
Dressler, A.,  1980, ApJS, 42, 565.

\bibitem[Dressler \& Gunn 1992]{dg92}
Dressler, A.\ \& Gunn, J.E., 1992, ApJS, 78, 1.

\bibitem[Dressler et al.\ 1994]{dobg94}
Dressler, A., Oemler, A., Butcher, H.\ \& Gunn, J.E., 1994, ApJ, 430, 107.

\bibitem[Dressler et al.\ 1997]{b97}
Dressler, A., Oemler, A.\ Jr, 
Couch, W.J., Smail, I., Ellis, R.S., Barger, A., Butcher, H.,
Poggianti, B.M.\ \& Sharples, R.M., 1997, ApJ, submitted.

\bibitem[Ebeling et al.\ 1996]{eb96}
Ebeling, H., Voges, W., Bohringer, H., Edge, A.C., Huchra, J.P.\
\& Briel, U.G., 1996, MNRAS, 281, 799.

\bibitem[Edge \& Stewart 1991]{ace91}
Edge, A.C.\ \& Stewart, G.C., 1991, MNRAS, 252, 428.

\bibitem[Edge et al.\ 1997a]{ace97a}
Edge, A.C., et al.\ 1997a, in prep.

\bibitem[Edge et al.\ 1997b]{ace97b}
Edge, A.C., Smail, I., Ellis, R.S.,
Allen, S.W., Fabian, A.C., Ebeling, H.\ \& Blandford, R.D., 1997b, in prep.

\bibitem[Ellis et al.\ 1997]{e97}
Ellis, R.S., Smail, I., 
Dressler, A., Couch, W.J., Oemler, A., Butcher, H.\ \& Sharples, R.M.,
1997, ApJ, 483, 582.

\bibitem[Faber et al.\ 1997]{f97}
Faber, S., et al.\ 1997, preprint.

\bibitem[Fusco-Femiano \& Hughes 1994]{fh94}
Fusco-Femiano, R.\ \& Hughes, J.P., 1994, ApJ, 429, 545.

\bibitem[Fioc \&  Rocca-Volmerange 1997]{frv97}
Fioc, M.\ \&  Rocca-Volmerange, B., 1997, in prep.

\bibitem[Gunn \& Gott 1972]{gg72}
Gunn, J.E.\ \& Gott, J.R., 1972, ApJ, 176, 1.

\bibitem[Hogg et al.\ 1997]{dwh97}
Hogg, D.W., Pahre, M.A., McCarthy, J.K., Cohen, J.G.,
Blandford, R.D., Smail, I.\ \& Soifer, B.T., 1997, MNRAS, in press.  (astro-ph/9702241).

\bibitem[Kodama \& Arimoto 1997]{ka97}
Kodama, T.\ \& Arimoto, N., 1997, A\&A, 320, 41. 

\bibitem[Koo et al 1997]{k97}
Koo, D.C., Guzm\'an, R., Gallego, J.\ \&
Wirth, G.D., 1997, ApJL, 478, 49. 

\bibitem[Kneib et al.\ 1997]{jpk97}
Kneib, J.P., Pell\'o, R.,
Mellier, Y., Soucail, G., Fort, B., Ellis, R.S., Arag\'on-Salamanca,
A., Smail, I.\ \& Miralda-Escud\'e, J., 1997, in prep.

\bibitem[Landolt 1992]{l92}
Landolt, A.U., 1992, AJ, 104, 340.

\bibitem[Larson, Tinsley \& Caldwell 1980]{ltc80}
Larson, R.B., Tinsley, B.M.\ \& Caldwell, C.N., 1980, ApJ, 237, 692.

\bibitem[Lea \& Henry 1988]{lh88}
Lea, S.M.\ \& Henry, J.P., 1988, ApJ, 332, 81.

\bibitem[Lilly et al.\ 1995]{cfrs1}
Lilly, S., Tresse, L., Hammer, F., 
Crampton, D.\ \& Le Fevre, O., 1995, ApJ, 455, 96.

\bibitem[Lilly et al.\ 1996]{cfrs2}
Lilly, S., Le Fevre, O., Hammer, F.\ \& Crampton, D., 1996, ApJ, 460, 1.

\bibitem[Lubin et al.\ 1997]{lml97}
Lubin, L.M., et al., 1997, in prep.

\bibitem[Moore et al.\ 1996]{bem96}
Moore, B., Katz, N., Lake, G., Dressler, A.\ \& Oemler, A., 1996,
Nature, 379, 613.

\bibitem[Moore et al.\ 1997]{bem97}
Moore, B., Lake, G.\ \&  Katz, N.,  1997, ApJ, submitted (astro-ph/9701211).

\bibitem[Mushotzky \& Scharf 1997]{ms}
Mushotzky, R.F.\ \& Scharf, C.A., 1997, ApJL, 482, 13.

\bibitem[O'Hely et al.\ 1997]{eoh}
O'Hely, E., et al.\ 1997, in prep.

\bibitem[Pierre et al.\ 1996]{p96}
Pierre, M., Le Borgne, J.F., Soucail, G.\ \& Kneib, J.-P., 1996, A\&A,
311, 413.

\bibitem[Smail et al.\ 1995a]{s95a}
Smail, I., Hogg, D.W., Yan, L.\ \& Cohen, J.L., 1995a, ApJ, 449, L105.

\bibitem[Smail et al.\ 1995b]{s95b}
Smail, I., Hogg, D.W., Blandford, R.D., 
Cohen, J.G., Edge, A.C.\ \& Djorgovski, S.G., 1995b, MNRAS, 277, 1.

\bibitem[Smail et al.\ 1997a]{s97a}
Smail, I.,
Dressler, A., Couch, W.J., Ellis, R.S., Oemler, A.,   
Butcher, H.\ \& Sharples, R.M., 1997a, ApJS, 110, 213.

\bibitem[Smail et al.\ 1997b]{s97b}
Smail, I.,  Ellis, R.S.,
Dressler, A., Couch, W.J., Oemler, A.,   
Butcher, H.\ \& Sharples, R.M., 1997b, ApJ, 479, 70.

\bibitem[Thompson \& Gregory 1993]{tg93}
Thompson, L.A.\ \& Gregory, S.A., 1993, AJ, 106, 2197.

\bibitem[van Dokkum \& Franx 1996]{vdf96}
van Dokkum,  P.G.\ \& Franx, M., 1996, MNRAS, 281, 985.

\bibitem[Wang \& Ulmer 1997]{gw97}
Wang, Q.D.\ \& Ulmer, M.P., 1997, preprint (astro-ph/9702069). (WU)

\bibitem[Wilson et al.\ 1997]{gw97}
Wilson, G., Smail, I., Ellis,
R.S.\ \& Couch, W.J., 1997, MNRAS, 284, 915.

\bibitem[Zabludoff et al.\ 1996]{z96}
Zabludoff, A.I., Zaritsky, D., Huan, L.,
Tucker, D., Hashimoto, Y., Shectman, S.A.,
Oemler, A.\ \& Kirshner, R.P., 1996, ApJ, 466, 104. 

\end{thebibliography}
\end{document}